\documentclass[12pt,preprint]{aastex}

\def\kms{\relax \ifmmode {\,\rm km\,s}^{-1}\else \,km\,s$^{-1}~$\fi}

\def\farcs{\hbox{$.\!\!^{\prime\prime}$}}

\def\arcmin{\hbox{$^\prime$}}
\def\arcsec{\hbox{$^{\prime\prime}$}}
\def\secd#1.#2{ #1\farcs#2 } 

\def\degree{\mbox{$^{\circ}$}}
\def\Mso{{M$_{\rm \odot}~$}}
\def\cm3{${\rm cm}^{-3}~$}
\def\etal{et al.$~$}

\def\mincir{\ \raise-2.truept\hbox{\rlap{\hbox{$\sim$}}\raise5.truept
    \hbox{$<$}\ }}
\def\magcir{\ \raise-2.truept\hbox{\rlap{\hbox{$\sim$}}\raise5.truept
    \hbox{$>$}\ }}

\def\mlr{\rm M$_\odot$~yr$^{-1}$}

\shorttitle{}
\shortauthors{E. Villaver, G. Garc\'{\i}a-Segura \& A. Manchado} 

\begin{document}

\title{The dynamical evolution of the circumstellar gas around low-and
 intermediate-mass stars I: the AGB}

\author{Eva Villaver\altaffilmark{1}}
\affil{Instituto de Astrof\'{\i}sica de Canarias,\\ C. V\'{\i}a L\'actea S/N, 
        E-38200 La Laguna, Tenerife, Spain.}
\altaffiltext{1}{Currently at Space Telescope Science Institute,
                 3700 San Martin Drive
                 Baltimore, MD 21218. \\e-mail:villaver@stsci.edu}

\author{Guillermo Garc\'{\i}a-Segura}

\affil{Instituto de Astronom\'{\i}a-UNAM \\ Apartado postal 877, Ensenada\\
       22800 Baja California, M\'exico\\e-mail: ggs@astrosen.unam.mx}
\author{Arturo Manchado}
\affil{Instituto de Astrof\'{\i}sica de Canarias,\\ c. Via Lactea S/N, 
        E-38200 La Laguna, Tenerife, Spain. \\e-mail:
        amt@ll.iac.es}

\begin{abstract}
We have investigated the dynamical interaction of low- and-intermediate
mass stars (from 1 to 5~M$_{\rm \odot}$) with their interstellar medium
(ISM). In this 
first paper, we examine the structures generated by the stellar winds
during the Asymptotic Giant Branch (AGB) phase, using a numerical code and
the wind history predicted by stellar evolution. The influence of the
external ISM is also taken into account. 

We find that the wind variations 
associated with the thermal pulses lead to                                    
the formation of transient shells
with an average lifetime of $\sim 20,000$ {\rm yr}, and consequently do not remain 
recorded in the density or velocity structure of the gas.
The formation
of shells that survive at the end of the AGB occurs via two main processes:
shocks between the shells formed by two consecutive enhancements of the
mass-loss or via continuous accumulation of the material 
ejected by the star in the interaction region with the ISM. 

Our models show that the mass of the circumstellar envelope increases
appreciably due to the ISM material swept up by the wind (up to $\sim$ 70 \%
for the 1 \Mso stellar model). We also point out the importance of 
the ISM on the deceleration and compression of the external shells. 
      
According to our simulations, large regions (up to 2.5 pc)
of neutral gas surrounding the molecular envelopes of AGB stars are
expected. These large regions 
of gas are formed from the mass-loss experienced by the star 
during the AGB evolution.
\end{abstract}


\section{Introduction}
After the completion of hydrogen and helium core burning, low-
and-intermediate mass stars ascend the Asymptotic Giant Branch (AGB) in the
HR diagram. At the end of the AGB phase, during the so-called thermal
pulsing AGB stage (TP-AGB), these stars experience high mass-loss
rates modulated by the occurrence of thermal pulses. A circumstellar envelope
(CSE) of gas and dust is formed which later on will be ionized by the
hot stellar remnant, forming a planetary nebula (PN). 

The mass-loss during the AGB
plays a decisive role in the subsequent evolution of the star. The star
detaches from the AGB when 
the stellar envelope mass falls below a critical value. Moreover, it is
during the AGB phase when
heavy elements which have been processed in the stellar interiors are
returned to the interstellar medium (ISM). The mass-loss itself also
determines 
the properties of the AGB population,
the distribution of white dwarf masses,
the maximum white dwarf progenitor mass (hence the minimum supernova
type II progenitor mass), and the PN structure. 

Mass-loss is
a crucial process in the formation of PNe and in the evolution of stars,
however, it cannot be calculated from first principles.
The origin of the high mass-loss rates (the so-called superwind (SW); 
Renzini 1981) needed to remove the stellar envelope
at the tip of the AGB is still a matter of controversy (Mowlavi 1998).
The scenario generally accepted nowadays involves two processes which work
together: shock waves caused by the stellar pulsation and the acceleration of
dust by radiation pressure. The stellar pulsation in AGB stars (Mira-type
stars) creates shock waves
which propagate through the stellar atmosphere. The dissipation of the
mechanical energy associated with these shocks leads to the levitation of the
upper layers of the atmosphere, where the gas becomes sufficiently cool (by
expansion and by dilution of the stellar radiation field) and dense
to allow heavy elements to condense into grains.
 As grains nucleate and grow they
experience the force exerted by the stellar radiation pressure and thus
are accelerated. The momentum coupling between gas and dust drives
the outflow. Pioneering work on dynamical calculations to drive the stellar
wind were first made by Wood (1979) and Bowen (1988).
However, it is possible that other mechanisms such as mass-loss induced by
rotation (Dorfi \& H\"ofner 1996), acoustic waves in
the stellar atmospheres (Pijpers \& Habing 1989), radiation pressure in
molecules (Maciel 1976), turbulent pressure (Jiang \& Huang 1997), Alfven
waves (Hartmann \& MacGregor 1980) and
magneto-hydrodynamic processes (Pascoli 1994), may also play an important
role in the mass-loss process. In addition, large-scale convection and
thermal expansion should also need to be taken into account in the mass-loss
mechanism. 

Dynamical model atmospheres are needed for the calculation of
mass-loss. However, problems arise when time-dependent dynamics (shock 
waves and
winds), radiation transfer (strong variable stellar radiation field) and dust
and molecular formation processes need to be
considered together. Computational limitations and our current
knowledge of the fundamental physical data make even the state-of-the-art
models far from ideal in several ways. For one, they are restricted to grey
radiative transfer. Winters \etal (1994) computed models that give mass-loss
rates as a function of the fundamental stellar parameters for stationary
atmospheres while 
dynamical model atmospheres have been computed by Bowen (1988),
Fleischer, Gauger \& Sedlmayr (1992), Arndt, Fleischer \& Sedlmayr (1997) and
H\"ofner \etal (1998). H\"ofner \etal (1998) demonstrated how changes in the
micro-physics, in particular, the use of more realistic opacities in the
atmosphere, may result in considerable lower mass-loss rates. The problem is
that obtaining reliable quantitative theoretical mass-loss rates from these
models is still not possible (Olofsson 1999).

Due to the inherent difficulties in the 
calculations of mass-loss, and in recovering the history
of mass-loss from observational studies, its exact evolution still remains
unknown. It is this basically unknown mass-loss history experienced by the
star during the AGB which determines the circumstellar structure at the 
end of the TP-AGB and, moreover, provides an
important clue for understanding PNe formation.
Stellar evolutionary models are able to 
provide the temporal behavior of the mass-loss during the AGB
and are available in the
literature (Vassiliadis \& Wood 1993; Bl\"ocker 1995; Schr\"oder \etal 1999). 
Although the mass-loss rates are not derived from first principles in these
models, and most
of them rely 
either on the dynamical model atmosphere calculations of Bowen (1988) and
Arndt \etal (1997), or on the semi-empirical mass-loss rates formula
derivations of Wood (1990), 
they do provide a unique opportunity to study the
extensive history of mass-loss on the AGB and beyond. 

Hydrodynamical
models of the evolution of a dusty circumstellar shell in the final stages of
the AGB evolution have been performed by Steffen, Szcerba \& Sch\"onberner
(1998) and Steffen \&  Sch\"onberner (2000). They used the long term
variations of the mass-loss for a 3 \Mso star during the AGB 
predicted by Bl\"ocker (1995) and a two component radiation
hydrodynamical code to drive the wind. Due to the inherent difficulties of
this kind of approach they did not solve the energy equation for the gas
component. The grid sizes they used do not allow to study the full mass-loss
process experienced by the star, they were limited to $1\times10^{18}$
{\rm pc} (Steffen \etal 1998) and $5\times10^{17}$ {\rm pc} (Steffen \&
Sch\"onberner 2000). The influence of the ISM could not be studied in these
models because the ISM is pushed out of the computational domain 
during the evolution .

Multiple dynamical interactions 
are expected through the wind which is strongly modulated 
by the thermal pulses. Our aim is to study the structures generated by
the full process of mass-loss during the AGB for different progenitors. For
this we have ensured that our grids are large enough to allow us to fully
follow the 
structures generated by the stellar ejecta. Thus the influence of the ISM on
the shell formation process can be addressed. 

We present a grid of models describing the dynamical gas
evolution during the AGB, encompassing the range
of initial masses of PN progenitors. We use the wind temporal behavior
predicted by Vassiliadis (1992) for stars with solar metallicity. In 
$\S2$ we describe our
numerical method, the initial and boundary
conditions adopted for the simulations, and the assumptions on which the
models are based. In $\S3$ we describe separately the results obtained
for each 
stellar mass. The discussion of the evolution
during the AGB with mass-loss is presented in $\S4$, together with a
comparative analysis for the different progenitors, the ISM influence and a
limited comparison of our results with the observations.

\section{The numerical method and computational details}
We have performed hydrodynamical simulations of the wind evolution
starting at distances from the star where the wind has reached
its terminal velocity. For the numerical simulations we have
used 
the multi-purpose fluid solver ZEUS-3D (version 3.4), developed by
M.L. Norman and the Laboratory for computational Astrophysics (LCA). This is a
finite-difference, Eulerian, fully explicit code which works
efficiently in one or two dimensions by choosing the appropriate symmetry
axis (for further details about the numerical methods used in the code see
Stone \& Norman 1992a; Stone \& Norman 1992b; Stone, Mihalas \& Norman 1992).

We have performed the simulations in 1D using spherical coordinates.
The radial sizes of the grids as well as the
cell sizes for each model are specified in Table~1. Column (1) gives the
(initial) mass of the models. The TP-AGB evolutionary
time and the time until 
the onset of the superwind are given in columns (2) and (3). We have adopted a
resolution of 1000 zones in the radial coordinate of the grid for all the
models, therefore, the cell size changes depending of the radial extension. 
We are interested in probing the effects of the mass-loss rate adopted
by the stellar models on the circumstellar structure, thus 
we should follow the whole stellar ejecta. To this end, the
radial extensions have been chosen to be large enough to avoid the mass-loss
flow out 
of the grid. Since the wind has reached its terminal velocity
at the innermost radial zone
in our simulation which lies at 8${\rm \times10^{16}}$ ${\rm cm}$ from the
central star, we do not need a two component (dust and gas) hydrodynamical
code to drive the wind. For the cooling curves given by Dalgarno \& McCray
(1972) and 
MacDonald \& Bailey (1981) a gas with solar composition has been used.

\subsection{The boundary conditions: the stellar evolutionary models}
The inner boundary condition for our simulations is determined by the way the
stellar wind flows from the star. The gas density at the inner boundary is
given by 
\begin{equation}
\rho(t)=\frac{\dot{M}(t)}{4\pi r^{2}v_{\infty}(t)}
\label{ro}
\end{equation}
where the temporal dependence of the mass-loss rate (\.M(t)) and terminal
wind velocity ($v_\infty$(t)) is given by the
stellar 
evolutionary models of Vassiliadis (1992) for stars with solar composition.
We have chosen these models because they provide with the temporal
evolution of \.M and $v_\infty$
during the AGB which is needed 
to study the long-term dynamical evolution of the gas. The 
parameterization made for the mass-loss and the  wind terminal velocity
is based on semi-empirical relations.

Vassiliadis \& Wood (1993) (hereafter VW93) computed the
evolution of stars from the Main Sequence 
through the Red Giant phase, to the end of the AGB.
In their formulation, the mass-loss during the AGB is derived from an
empirical formula computed from observations of Mira variables and pulsating
OH/IR stars in the Galaxy and the Large Magellanic Cloud. Based on the
empirical relation between the  
stellar pulsation and the period given by Wood
(1990), they established that the mass-loss rate is a function of the
pulsational 
period. They then derive \.M 
from fundamental physical parameters such as the mass and the radius of the
star by assuming that Miras, and the AGB stars surrounded by dust are
pulsating in the fundamental mode. They consider two distinct
phases for mass-loss: for stellar
masses below 2.5~M$_{\rm \odot}$, and for periods less than $\sim$500 days the
mass-loss increases exponentially with the period (Mira phase), and for stars
with mass greater than 2.5~\Mso they delay the onset of the superwind to
periods of $\sim$ 750 days. For periods beyond $\sim$ 500 days, the mass-loss
rate is essentially constant, and lies within a factor of $\sim$ 2 of the
value given by the radiation pressure limit.
The terminal wind velocity ($v_\infty$) is computed assuming the empirical
relation for 
the pulsation period derived by Wood (1990), and it is restricted to lie in
the range 3.0-15.0 \kms. 

We take the temporal evolution of the wind parameters for stellar masses 1,
1.5, 2, 2.5, 3.5, and 5~\Mso with solar metallicity (Z = 0.016) 
from the stellar evolutionary models computed by Vassiliadis (1992), and set
them in the 
five innermost zones of the grid for the AGB evolution of the star. In
Fig.~1 we show the evolution of the mass-loss and terminal
wind velocity used as the input for our models. We have labeled the
main wind modulations (e.g {\it I}, {\it II}) in order to allow a better
description through the text of the features they generate in the gas. The
wind temperature is assumed to be the effective temperature of the star. A
free-streaming boundary condition was set at the outer radial direction. 

Fig.~1 shows that both \.M and $v_{\infty}$ suffer from strong
modulations. The wind density is a bimodal function of these parameters, and
therefore whenever one changes so does the density. The density will be
softened when \.M and $v_{\infty}$ change in such a way that they neutralize
each others effect.

\subsection{The initial conditions}
The mass-loss experienced by low-and intermediate-mass stars
during the main sequence and red giant phases ($\sim 10^{-8}$ \mlr) is
negligible compare to the  
mass-loss during the AGB (up to $\sim 10^{-4}$ \mlr). The local ISM is not
significantly modified by the wind  
of previous evolutionary stages, due to the low mass-loss rates and velocities
involved. Therefore, in our study, we have considered an homogeneous ISM
as the initial condition.

The main component of the ISM is hydrogen in both molecular and
atomic form, and dust, although the amount of mass residing in the form of
dust is much smaller than that residing in gas (estimates of the dust-to gas
mass ratio are $\sim 3\times10^{-3}$; Burton, Elmegreen \& Genzel 1992). 
Although molecular hydrogen dominates over all gaseous material in the
temperature range from a few to about 30~K, it is confined to clouds (small,
massive and compact regions) which have an annular distribution in the inner
Galaxy (from 3 to 8 kpc). The 
diffuse ISM is thought to consist of four major 
components: the cold neutral medium (CNM), the warm neutral medium (WNM),
the warm ionized medium (WIM), and the hot ionized medium (HIM). Neutral
atomic hydrogen (HI) is the main observed constituent of the ISM; estimates
of its filling factor ranges from 20\% to 90\% (Burton 1988). The WNM is
the ISM component which occupies most of the space 
(Kulkarni \& Heiles 1988) in an ``intercloud medium'' or surrounding the cold
clouds. Since information about the WNM mainly comes from 21 cm line emission
data, direct measurements of
temperatures (estimated
to be in the range 5000~K to 8000~K) or densities (estimated to be between 0.1
and 1 $cm^{-3}$) are difficult to obtain. Therefore, the WNM is the
least well understood component of the ISM. A typical density of 0.4 $cm^{-3}$
characterizes much of the ISM (Burton 1988). 

We assume that the ISM has the
characteristics of the WNM because it is the main observed constituent of the 
ISM. We have considered that the ISM pressure is simply the
standard gas kinetic pressure, $P = nKT$ (where $n$ is the number of
particles per unit volume, $K$ is the Boltzman's constant and $T$ is the
gas temperature). Spitzer (1978) demonstrated that the contribution of
cosmic  
rays, and magnetic pressures can be comparable to the gas kinetic
pressure. As we
have not included magnetic fields or turbulent motions in our simulations, we 
therefore, use large densities (1 $cm^{-3}$ instead of the typical 0.4
$cm^{-3}$) to compensate for the cosmic rays, magnetic ($B^2/8\pi$)
and turbulent pressure components (assuming all of them contribute to an
isotropic pressure).  

In our models, at time zero the grid is filled homogeneously with neutral
atomic hydrogen 
with a density of 1 ${\rm cm^{-3}}$, a temperature of 6000~K, and 
zero macroscopic velocity. Hence, we have to
consider wind propagation in an 
homogeneous (cloudless), stationary ($v = 0$) and warm ($T=6000~{\rm K}$)
ISM (the isothermal sound speed is $c_{\rm s} = 7~$\kms). The
interaction of the stellar wind with the ISM 
influences the subsequent evolution of the flow. 
The adopted ISM
pressure of 6000 ${\rm K~cm^{-3}}$ lies in the range of the ISM pressure which
allows pressure equilibrium between the different ISM components. In
particular, 
thermal equilibrium between the WNM and the CNM is only possible over a range
of pressure between $\sim$2500 and $\sim$6500 (in units of the the Boltzman's
constant). The gas evolution considering an ISM with pressure
ten times lower is also presented in $\S4$.

In order to keep the ISM at its original temperature, no source of gas heating
is included. The temperature of the ISM decreases due to radiative cooling
during the evolution a negligible amount ($\sim$ 500~K). 

\section{Results}
The beginning of the evolution is the same for all the stellar models. A wind
with a constant mass-loss rate ($\sim 10^{-8}$ \mlr) and velocity ($\sim$3
\kms) starts to interact with the ISM. The duration of this 
stage of constant mass-loss rate at constant velocity
depends on the stellar mass (see third column in Table~1)\footnote{We
  considered that the so-called {\it superwind} phase starts when the 
  mass-loss increases to values greater than $2\times 10^{-7}$ {\rm
        M$_\odot$~yr$^{-1}$}.}.
The result is a compression wave propagating through the ISM at the speed of 
sound (7 \kms for $T_{ISM} = 6000~{\rm K}$). It should be noted that since the
wind 
temperature is assumed to be the $T_{eff}$ of the star ($\sim$ 2500 {\rm K})
a supersonic interaction is expect to occur for lower velocity
values. Therefore,
despite the low values 
reached by the wind parameters at the wind modulations labeled as {\it I} in
Fig.~1 for the 1~\Mso model, and at {\it I} and {\it  II} in the 1.5, 2 and
2.5~\Mso models, all of them are able to develop shock waves initially.
The subsequent evolution of the flow is related to the wind history of each
stellar model and are described below. 

The minimum and
maximum values reached 
by the wind temperature during the evolution are 2630~{\rm K} and 3160~{\rm 
K} respectively. The relative importance of the thermal pressure of the wind
in 
the shell morphology, although small at the end of the evolution still
affects the final size of the shells.

In the following, we present the results of the hydrodynamical simulations 
of the evolution of the stellar wind ejected during the
TP-AGB for each stellar mass considered. We
have selected the most representative outputs in order to describe the
evolution, avoiding the transient features appearing in the flow that rather
complicates the description.

\subsection{The 1~\Mso star case}
The dynamical evolution of the gas is described by relating it with the
evolution of the central star wind. As explained in $\S2$, the time-dependent
values of the mass-loss rate, velocity and wind 
temperature are used as the inner boundary condition (see top of Fig.~1
for the 1~\Mso values). The mass-loss 
reaches the radiation pressure limit (i.e the maximum mass-loss
value) only during the last four thermal pulses. Between the pulses the
mass-loss rate and velocity  
decreases, due to the extended luminosity dips which characterize the thermal
pulses for low-mass stars.

In Fig.~2, we have summarized the AGB evolution of the flow by
plotting five radial density and velocity profiles at
different stages during the evolution.
From top to bottom, the plots have been selected at 78, 84, 90,
97, and 99 per cent of the total AGB time, which in
Fig.~1 corresponds to the gap between pulse {\it II} and {\it III}, to the
beginning of 
{\it III}, to 20,000 {\rm yr} before the end of {\it III}, to
the gap between {\it III} and {\it IV}, and to the middle of {\it IV}
respectively. We have also marked each density peak 
and its corresponding velocity.

The main major changes in the gas structure appear as a consequence of
the wind modulations associated with pulse {\it II}
in Fig.~1. The increase in mass-loss rate
and velocity during this pulse causes the pile up of material
identified as feature {\it\bf a}\footnote{Hereafter, shell features are
 singly called with the letters they are referred to in the figures.},
(i.e. a shock structure). At the end 
of {\it II}, the wind velocity 
drops to nearly half of its value during the pulse, while the mass-loss rate
remains constant. A drop in velocity at constant mass-loss rate gives
rise to an increase in
density in the inner parts of the grid. Denser material from the inner
regions moves outwards at lower velocities than the material ejected
previously, hence forming a gap in the density  
structure. The shell identified as {\it\bf b} is been formed by this
process. When the high mass-loss rate during pulse {\it II} ceases,
{\it\bf c} is 
formed. In the top panel of 
Fig.~2 we show the density structure $\sim$ 7,400 {\rm yr} after
the formation of {\it \bf c}. Both {\it \bf b} and {\it \bf c} are 
accelerated by pressure gradients, whilst {\it \bf a} is decelerated by
the ISM.   

About 32,000 {\rm yr} later (second panel from top) only one density peak,
{\it\bf ab}, formed when {\it\bf b} reaches {\it\bf a} and they merge, is
visible. Due to velocity gradients  
{\it\bf c} has broadened and has disappeared. At the time shown, 
pulse {\it III} has already started, and therefore 
the density and velocity at the inner parts of the grid have increased (see
also the third plot from the top 32,000 {\rm yr} later). 
Due to pulse {\it III}, denser material moving at a higher velocity reaches
the previously  ejected material. A shock develops in the region of
interaction between the mass lost during pulse {\it II} and that occurring 
during {\it III}. This shock region will give rise to the formation of the
shell identified as 
{\it\bf x} (see  fourth plot from top). Shells {\it\bf d} and {\it\bf e} are
formed by the same process which formed shells {\it\bf b} and {\it\bf c},
i.e. a gradual velocity drop at the end of {\it III} (forming
{\it\bf d}) and the subsequent drop in mass-loss rate which forms {\it\bf e}.

The gas structure
7,600 {\rm yr} later is shown at the bottom panels of Fig.~2. In this
case, it is {\it\bf d}, formed by the velocity 
drop at the end of the pulse, which broadens and
disappears, whereas {\it\bf e} last for a longer time because it is not 
captured by denser material leaving the star. The gas density reaches
higher values during pulse {\it IV} due to the small increase in
the wind velocity. 

In Fig.~3 we show the temporal behavior in
terms of radius, velocity and density of the shells formed in the
CSE during the TP-AGB evolution for a star with 1
\Mso. Note that only 
shell {\it\bf x}, formed as a consequence of a shock, and shell {\it\bf a},
due to the capture of {\it\bf b}, increase in density. Note that all the
shells are decelerated during the evolution.

\subsection{The 1.5~\Mso star case} 
The temporal behavior of the wind parameters used as the inner
boundary condition for the study of the 1.5\Mso model are shown in the second
panel of Fig.~1. 
The circumstellar gas evolution during the TP-AGB is plotted in
Fig.~4. The five outputs shown in Fig.~4 have been selected
at times which corresponds in Fig.~1 to  
12,300 {\rm yr} after the end of {\it III}, 
95,600 {\rm yr} since the begining of {\it IV},
15,000 {\rm yr} before and after the end of {\it IV} and 
43,000 {\rm yr} since the beginning of {\it V}. The percentage of the
total AGB time-life at which the density and velocity profiles are shown is
specified in the upper right corner of the density plots.

Despite the different wind history experienced by stars with 1 
and 1.5 $M_{\odot}$, the general behavior of the circumstellar gas is very
similar. The density peaks {\it\bf a}, and {\it\bf b} seen in the top panel
of Fig.~4 are formed by the mass
accumulated during {\it III}, and by the reduction of mass-loss and
velocity in the aftermath of this pulse respectively.
Pressure gradients are responsible for the acceleration of {\it\bf b},
while {\it\bf a} is decelerated by the pressure exerted by the
ISM gas which lies ahead. Later on ($\sim$30,000 {\rm yr}), these two shells
merge and {\it\bf ab} is then formed. 

The density and velocity
profiles in the third plot of Fig.~4 have been disrupted by the higher
mass-loss and 
velocity wind  
arising from the star, i.e. pulse {\it IV} has begun. 
As in the 1 \Mso~case, a shock develops at the position where
the mass lost by {\it IV} encounters the previously
ejected material which has been slowed down during the evolution. 
Later on, {\it\bf x} will appear at this position. Shell {\it\bf c} is
formed when {\it IV} ends $30,000$ {\rm yr} later (fourth plot from top).
This shell is 
smoothed rapidly due the large velocity gradients. The transient
feature labeled {\it\bf d} is formed by the mass-loss rate
and velocity increases at {\it V}. We can also see how, as a consequence of
the shock, matter in {\it\bf x} is compressed. 

The shell {\it\bf x1}, visible in the bottom plot of Fig.~4, 
is formed in the interaction region between the material which has been 
ejected during pulse {\it V} and the one ejected during pulse {\it IV}.

\subsection{The 2~\Mso star case}
The evolution of the mass lost by a star with a main
sequence mass of 2~\Mso is shown at ten different times in 
Figs.~5 and 6.  
The gas density and velocity radial profiles have been selected at times which
correspond in Fig.~1 to 23,000 {\rm yr} after
the beginning of {\it II}, 4,000 {\rm yr} after the end of {\it II}, 16,000
{\rm yr} after the end of {\it II}, 3,000 {\rm yr} after the end of {\it III}
and 6,000 {\rm yr} after the beginning of {\it IV}.

Both {\it\bf a} and {\it\bf b} in Fig.~5 are caused by
{\it II}, the former by the pile up of material during this pulse, and the
latter when the pulse has finished.  They
merge $\sim$ 18,000 {\rm yr} later, 
forming {\it\bf ab}. Shell {\it\bf ab} propagates
outwards, and finally disappears only slightly modifying the surrounding
gas. Shells {\it\bf c} and {\it\bf d} have the same formation process as
shells {\it\bf a} and {\it\bf b} mentioned previously, but in this case their
origin is associated with {\it III}. Again, because {\it\bf c} is being
decelerated, {\it\bf d} catches it up in $\sim 3\times10^4$ {\rm yr} and
hence {\it\bf cd} appears.

In Fig.~6 we have shown the gas density and velocity
evolution at times which correspond in 
Fig.~1 to 4,500 and 3,200 {\rm yr} before and after the end of
{\it IV} respectively and 4,500, 16,500 and 71,300 {\rm yr} after the beginning of
{\it V}. At the top of Fig.~6 we have marked the density
peak {\it\bf e} which is formed by {\it IV}. A strong shock develops 
in the density valley formed between 
{\it\bf cd} and {\it\bf e}, leading to the formation of {\it\bf x}. 
When pulse {\it IV} finishes, shell {\it\bf f} appears. 
With increasing time, all the density peaks
({\it\bf cd}, {\it\bf e}, {\it\bf f}) disappear because they feed {\it\bf x}.
Only {\it\bf x} remains
as an identifiable shell until the end of the AGB.
Pulse {\it V} is responsible for the increase labeled as {\it\bf g} 
which completely disappears 4,400 {\rm yr} after when the density falls 
as a power law of the type $r^{-2}$ law. This is what is expected 
for a constant mass-loss rate at a steady wind velocity.
The disruption of the density structure seen in the bottom plot of
Fig.~6 is caused by the velocity decrease produced during 
the middle of pulse {\it V}. 

\subsection{The 2.5~\Mso star case} 
Of all the cases considered this model has the longest evolution during the
AGB. The wind parameters undergo three main
modulations in mass-loss rate and velocity, which occur during the 
last 300,000 {\rm yr} (see Fig.~1).  
The evolution of the gas is displayed at five snapshots during the TP-AGB in
Fig.~7. The times selected have a relation to the stellar evolution
as follows: the top plot corresponds to 6,500 {\rm yr} after the end of
{\it II}, the second and third plots have been selected at 39,000 and 8,000
{\rm yr} after the begining and end of {\it III} respectively, and the fourth and
fifth plots at 45,500 and 12,300 {\rm yr} after the begining of {\it IV}. 

The small modulations in velocity experienced by the wind before pulse 
{\it II} do not have any effect on the gas structure. The
first significant effect arises as a consequence of {\it II} when a high
density shell is formed. Due to its high thermal pressure this shell 
expands 
splitting the density and velocity into two peaks. This peculiar two peak
structure lasts for around 22,000 {\rm yr}, disappearing as the gas propagates
outwards, resulting in the formation of {\it\bf a} (see top plot 
in Fig.~7). Shells {\it\bf b} and {\it\bf b1} are formed by the gradual end
of {\it II}: first, both the wind velocity and the mass-loss rate
are reduced by factors of 3 and 5 respectively ({\it\bf b} is formed), 
and second, the mass-loss rate drops again a factor of 1.5, meanwhile the wind
velocity 
grows in {\it III} and {\it\bf b1} is formed. Shell {\it\bf b1} survives
for only a short time (8,000 yr), it is accelerated by the incoming
wind and quickly merges with {\it\bf b}.

A density peak around 0.1 pc appears in the second plot of Fig.~7 due to the
fast increase in mass-loss and velocity experienced
by the wind towards the end of {\it III}, {\it\bf c} is
formed when {\it III} ends. A shock where the gas is compressed
({\it\bf x} in the fourth plot) is produced in the region of interaction
between structures {\it\bf ab} and {\it\bf c}.

All the modulations produced during {\it IV} give rise 
to the formation of transient density peaks, such as {\it\bf d}, which are
smoothed out during the evolution of the flow. Only {\it\bf x} remains at the
end of this stage (see bottom plot of Fig.~7). 

\subsection{The 3.5~\Mso star case} 
Both the wind velocity and mass-loss rate for the 3.5~\Mso model grow
continuously during the TP-AGB, suffering only from small sudden drops in
intensity (see Fig.~1). 
 
The gas evolution is shown in Fig.~8, where the plots have been selected in
order to show the formation of structures that appear after the three major
changes in the wind parameters.  
Shell {\it\bf a} is formed at 220,000 {\rm yr} 
when the mass-loss rate reaches a value of $5\times10^{-7}$ \mlr
and the wind velocity is 8 \kms.
This shell is fed continuously by the stellar wind.
There are three important drops in the wind velocity and mass-loss rate
that give rise to the formation of the transient features labeled
{\it\bf b}, {\it\bf c} and {\it\bf d} respectively.  
Only one shell ({\it\bf abcd}) is visible in the bottom plot of
Fig.~8 which is the result of the accumulation of matter
during the AGB evolution of the star.

\subsection{The 5~\Mso star case} 
The relatively simple behavior of the input wind parameters for the star with
5 \Mso (without abrupt changes in mass-loss or 
velocity, see Fig.~1) is translated directly into a simple evolution of the
circumstellar gas. 
As expected, a $\rho \propto r^{-2}$ law
characterizes the radial density profile of the gas up to the position of
an external shell. This cold external shell is formed and fed with the high
mass-loss rates from the star and
propagates outwards with little disruption. In Fig.~9
the evolution of this shell since its formation is shown at five different
times. The top plot in Fig.~9 corresponds to
80,000 {\rm yr} after the beginning of the AGB evolution and the following 
plots have been chosen at time intervals of $1\times10^4$, $5.4\times10^4$,
$5.5\times10^4$, and $5.2\times10^4$ {\rm yr} respectively.

\section{Discussion}
In the previous section, we summarized the circumstellar gas evolution
during the TP-AGB phase and related it with the stellar wind behavior. The 
dynamical evolution of the circumstellar gas is highly non-linear, and
although the stellar 
wind on the TP-AGB experiences huge modulations,
the circumstellar gas structure does not retain this information. The 
subsequently formed shells are either accelerated during the 
evolution, or decelerated by the ISM, most
of them disappearing when they merge with other shells or when they
reach pressure equilibrium with the surroundings. 
The shells
formed by a shock in the interaction region between two subsequents
enhancements of mass-loss and velocity are the most stable structures, and
remain visible at the end of the AGB phase. 

In table~2 we have summarized, for each of the shells marked in Figs.~2, 4, 5,
6, 7, 8 and 9 the radius (determined at the position of the density peak),
density and velocity for the times shown in 
the figures. The formation process of the shell and its life-time ($L_t$)
are also given, where ``$\infty$'' means that the shell remains until the end
of the TP-AGB evolution. The largest radius corresponds to the shells with
the longest lifetime. It should be noted that although shells {\it\bf ab} for
both the 1 and 1.5~\Mso models have the largest radii, they are not the
main observable feature. Shell {\it\bf x} is the main observable feature due
to its highest density. 
The radii show a range between $\sim$0.07 {\rm pc}, for shells that have just
been 
formed, up to $\sim$1.8 {\rm pc} for evolved shells at the end of the AGB.
The shells velocities range between 2 \kms up to 19 \kms.
Although the maximum wind velocity was 15 \kms we can see how many of these
shells have been accelerated by thermal pressure. 

In Fig.~10 we show the circumstellar gas density and velocity structure at
the end of the AGB for all the initial masses
considered. For the 1 \Mso model, the bottom
panel of Fig.~2 was selected very close to the end of the
AGB stage, and therefore the gas radial profiles are very similar to
the ones shown in Fig.~10. In the case of the 1.5 \Mso~model only one external
thick shell remains visible which is formed
when shell {\it\bf x} merges with shell {\it\bf ab}.
A double-peaked structure can be seen in the outermost shell
of the 2~\Mso model where the outermost peak correspond to shell {\it\bf x}
(see Fig.~6) and the innermost 
peak is produced by a shock region which develops when the matter ejected
later on from the star reaches shell {\it\bf x}. Therefore this double-peaked
structure is not caused by an expansion due to a pressure difference. 
The density structure of the
rest of the models has
not changed significantly since the times shown in
Figs.~7, 8, and 9. 

It has been widely assumed in the literature that a law of the type
$\rho~\propto~r^{-2}$ describes the gas structure resulted from the AGB
evolution (e.g. Okorokov et 
al. 1985; Schmidt-Voigt \& K\"oppen 1987a, 1987b;  Marten \& Sch\"onberner
1991; Mellema 1994; Frank \& Mellema 1994).
In order to compare our density structures with a constant mass-loss rate at
constant velocity density law we have overplotted in Fig.~10 a density law of 
the form $\log \rho =-2~\log(r)+C$ where $C=\log(\dot{M}/4~\pi~v_{\infty})$.
To compute $C$ we have used the values of 
\.M and $v_{\infty}$ reached by the wind at the end of the AGB. The 1~\Mso
model is different to the others in the sense that at the 
end of its evolution, when pulse {\it IV} occurs, the velocity
remains very low (in contrast to what happens during previous 
pulses). This results in a huge increase in density 
in the inner parts of the grid (see Fig.~10). Two 
fits have been made for this model; the dotted line represents the fit 
for the real values of \.M and $v_{\infty}$ reached by the wind at the end of 
AGB phase, the dashed-dotted line is for a wind with a terminal
velocity of 15 \kms and the \.M experienced by the star.
The values for the
constant $C$ adopted to fit the constant mass-loss rate, constant velocity law
are plotted in Fig.~11. The relationship of $C$ with stellar mass can be
described by a linear fit of the form
$C=-23.79+0.16 M (M_{\odot})$ (for $M~>~1$ \Mso), which is shown in
Fig.~11. The same values of $C$ which fit the density profiles are obtained
when other ISM densities are used in the simulations.

The common
factor in all these models, with the 
exception of the 1~M$_{\rm \odot}$, is that the complicated wind history
experienced 
during the AGB results in the formation of simple structures. The density
distribution can be described by a $r^{-2}$ density law except for the
outermost shell. The gas structure produced by the
evolution does not retain the
information about the wind history experienced by the stars. The low-mass
stars 
undergo four or five significant modulations in the wind parameters during
the AGB.  
However, except for the 1~\Mso star model, only one shell
is visible in the density structure. 
Therefore, according to our models, is not possible to recover the history of
mass-loss experienced by the star from the density structure at the end of
the AGB. Moreover, the prominent peaks are not signatures of mass-loss rate
enhancements 
but are the consequence of the development of shocks in the interaction
region between two important mass-loss events. Another important fact is that
the shells disperse on timescales of the order 
of 20,000 {\rm yr} (much shorter than the characteristic AGB evolutionary
time), independent of the formation process. The only exceptions are the
shells 
formed by shocks {\it\bf x} which remain as stable structures in the density
profiles 
at the end of the AGB.

According to the formation
process of the outer shell we can
classify the models into two mass groups. Firstly, the group formed by
the 
stellar masses 1, 1.5, 2 and 2.5 \Mso for which the outer shell is
formed by a shock in the interaction region between two subsequents
mass-loss rate events, and secondly, the group formed by stellar masses
3.5 and 5~M$_{\rm \odot}$, for which the external shell is formed by the
accumulation of 
mass during the course of the evolution. Thus, low- and
intermediate-mass stars have different outer shell formation processes.

The stellar models with masses 2, 2.5, 3.5 and 5~\Mso share similar radial
velocities profiles. All of them are
characterized by a continuous 
increase (with a small slope) up to the position of the external shell,
where the velocity reaches its highest value of $\sim 20$ \kms. Moving
outwards, the sharp fall off in the velocity is interrupted by the presence
of a small plateau at the position of the outermost density peak
(i.e. external shell).  

\subsection{The role of the ISM}
A novel aspect that
makes our study different from previous ones 
(e.g. Steffen \etal 1998, Steffen \& Sch\"onberner 2000) is the use of grids
large enough to prevent 
the flowing out of stellar ejecta. Thus, we can quantify the effect of
the ISM on the gas structure, since we have ensured that the ISM is not
completely pushed out of the computational domain. When the wind flows from
the star it encounters a non-negligible pressure provided by the ISM. The ISM
is pushed and 
compressed while the expansion of the stellar wind is slowed 
down. However, we should keep in mind that the pressure exerted by the ISM
can change radically depending
on the position of the star within the Galaxy. In particular, the ISM density
fall off exponentially in the direction perpendicular to the 
Galactic plane from a value of $\sim 2~{\rm cm^{-3}}$ (Spitzer 1978) and a
scale height of 100 {\rm pc} (Mihalas \& Binney 1981). 

In order to consider the evolution 
of stars at higher Galactic latitude, 
we have also performed the simulations using an ISM ten times
less dense than the one used in the models described in the previous sections.
This gives us an external pressure ten
times smaller, because only the thermal pressure component is considered in
this study.
The gas structure at the end of the AGB as a 
result of the evolution through this low density ISM is shown in 
Fig.~12 for all the initial stellar masses considered. 
The dotted line represents a fit using a $r^{-2}$ density law.
A first glance at Fig.~12 reveals
that the external shell has lost density contrast compared to the models
evolving though a medium with higher pressure.

\subsubsection{The mass of the circumstellar envelope} 
The dynamical
interaction of the stellar wind with its environment is important to 
understand the mixing of the ejecta with 
the ISM (fundamental for chemical abundance studies). As the stellar ejecta
evolves it sweeps up ISM material that has not been enriched in the stellar
interiors. In order to address the relative importance of this mixing effect 
we compute the fraction of the mass contained in the circumstellar
shell that belongs to the ISM. This can be done in our case since we
know the amount of mass returned to the ISM by the star.
We integrate the total amount of mass in the grid which is above
the ISM density at the end of the AGB for each stellar mass. This is the mass
we might expect to observe above the background level.
This is done for the two
ISM densities considered and is shown as a function of the stellar
mass in the left panel of
Fig.~13. The long-dashed and dotted
lines represent the integrated mass for ISM  
densities of 1 and 0.1 $cm^{-3}$ respectively.
The solid line represents the total mass lost by the star during the 
course of the AGB evolution. The integrated mass in Fig.~13
approaches the total mass lost by the star when the density of
the ISM decreases. 

We compute the fraction of the mass residing in the CSE that belongs to
the ISM by subtracting from the integrated mass above the ISM level
the mass lost by the star, and then dividing by the initial stellar
mass. This is 
shown in the right panel of Fig.~13 for the two ISM density values considered.
As expected, the percentage of ISM material in the CSE decreases with the ISM
density. However, even for a low density medium 
this fraction is not a negligible amount. For a star with
1 \Mso evolving through a medium with density of 1 ${\rm cm^{-3}}$,
approximately 70 \% of the matter in the CSE belongs to the ISM. This amount
decreases to $\sim$ 20 \% when  
the medium is 10 times less dense. The swept up ISM matter 
also decreases with increasing main sequence stellar mass.
If a low mass progenitor evolves in a high density medium,
up to $\sim 70\%$ of the measured mass is ISM matter which has not been
processed in the stellar interior. This can have important consequences for
abundance analysis, since the observed gas is mixed with non
chemically enriched ISM matter. 

The presence of this
non negligible amount of matter from the ISM in the CSE leads to another
important problem. The mass of the progenitor derived from
observations of the CSE is based on the assumption that all the observed
matter has been lost by the star. We have seen that the mass in the CSE
increases due to the swept up ISM material, and this effect
should be taken into account in order to not overestimate the mass
of the progenitor from observations of CSE.

\subsubsection{The ISM pressure} 
The size of the CSE reached at the end of the TP-AGB
depends on the external pressure, the momentum
of the wind, and on the time they are allowed
to evolve. In order to quantify the relative importance of all these
quantities, we have plotted in the left panel of Fig.~14 the radius of the
CSE at the end of the TP-AGB as a function
of the initial stellar mass for the two ISM densities considered
($1~{\rm cm^{-3}}$ long-dashed line and $0.1~{\rm cm^{-3}}$ dotted line). The
radius has 
been determined as the largest distance from the star where the ISM has been
modified, we can see that the radius increases by a factor of $\sim
1.5$ when 
the ISM density decreases by a factor of ten.
In the right panel of Fig.~14, we show the TP-AGB evolutionary time versus the
stellar mass (solid line), and the kinematical ages of the envelope for
the two cases 
considered of ISM densities, $1~{\rm cm^{-3}}$ and $0.1~{\rm cm^{-3}}$
(long-dashed 
line and dotted line respectively).
The kinematic age of the 
envelope has been computed by using the radii plotted in the left panel of
Fig.~14 and assuming a constant outflow velocity of 15 \kms (a common value
usually reported in the literature; i.e. Young, Phillips \& Knapp 1993a). 

We can see from the right panel of Fig.~14 that 
the kinematic ages computed are almost 
the same for all the stellar masses, even when the differences in the
evolutionary times are as large as a factor of ten. A possible 
consequence of this results is that the CSE kinematical ages are not ideal to
constrain the evolutionary age of the star, thus the progenitor mass. 
The largest radius is reached by the star with initial mass 2.5~M$_{\rm
  \odot}$, i.e. the star that has the longest evolutionary time. The secondary
maximum in radius (for the 1.5~\Mso model) is related with the total 
time the wind is supersonic. Pressure equilibrium with the ISM is not 
reached in any of the two cases of ISM pressures considered. 

In  
Table~3 we have listed the sizes, densities and velocities of the
circumstellar shells at the end of the AGB, where cases A and B represent the 
two ISM densities considered, 1 and 0.1 ${\rm cm^{-3}}$ respectively. These
values 
have been determined at the position of the outermost density maximum, and
are aimed to illustrate the effect of the ISM pressure on the CSE
properties.   
Both values considered for the thermal pressures of the ISM, 6,000
and 600 ${\rm cm}^{-3}$ K (where K is the Boltzman's constant)
are able to confine the shells (note the 
density enhancement at the border of the density profiles). An ISM
with a thermal pressure ten times smaller (case B in Table~3)
results in external shells with average radii 1.1 times larger, shell
densities 0.2 times smaller and expansion velocities which are 1.5 times
faster. Moreover, the kinematical ages derived for the ISM with less
pressure are on average 0.8 times smaller. 

The net effect the ISM has on the formation of the CSE is to decrease its
size, stall its expansion, and increase its 
mass. Moreover, and even more important, a non-negligible amount of ISM
material is mixed in the CSE with the chemical enriched material ejected from
the star. Thus, the effect of the ISM is very important
and should not be neglected in further studies.

\subsection{Comparison with the observations}
The CSE can be observed at infrared wavelengths, where the emission comes
mainly 
from the heated dust, and in the radio where it is produced mainly by
molecules such us CO and OH. Because the molecular line emission arises from
photodissociation, 
its extent is limited to $\sim 10^{16}$ {\rm cm} for thin shells, although it
can be 
larger by up to few times $10^{17}$ {\rm cm} for thicker envelopes (Mamon
\etal 
1988; Letzelter \etal 1987). Therefore, using observations of the molecular
emission 
from CSE, one can only probe the  recent mass-loss history experienced by the
star (Olofsson \etal 1993). From an observational point of view,
tracing the mass-loss history as far 
back as $\sim 10^5$ {\rm yr} needs to be done at infrared wavelengths, but
the dust has low emissivity at the temperatures and distances involved.

According to
our models, large CSE are formed around AGB stars, with shells sizes
ranging from 1.7 to 2.6 pc (see Table~3) depending on the 
progenitor mass and on the ISM pressure. 

A large CSE surrounding the oxygen-rich `red giant' W Hydrae was discovered
with IRAS observations by Hawkins (1990). Emission was detected out to 40
\arcmin~(up to 35 \arcmin at the 6 sigma level) which gives 
a diameter larger than 1.6 pc, assuming a distance of 135 pc.
W Hydrae is one of the nearest red giant star
and it is found at high galactic latitude in an area of 
low infrared cirrus emission that might otherwise confuse its
surrounding extended emission. 
Gillett \etal (1986) reported the observation of another large CSE around
R Coronae borealis. It has an extended shell with dimensions $r_{\rm min}=0.65
$ {\rm pc} and $r_{\rm max}=4.3$ {\rm pc} (assuming a distance of 1.6
kpc). The galactic 
latitude of R CrB is 51 \degree, which places it at about 1.24 kpc above
the Galactic plane. 

Of the 76 CSE resolved in the 60 $\mu m$ band in the IRAS survey of
Young, Phillips \& Knapp 1993b), 26 have outer
radii larger than 0.7 pc, and 12 of these have radius between 1.2 and 2.5 pc,
in excellent agreement with our predictions. The shell average radius for the
whole Young \etal (1993a) sample is 0.74 pc. Three carbon
stars W Pic, RY Dra and R CrB were found to have shells with radii
of 3 pc or more.  

Recently, Speck, Meixner \& Knapp (2000) have found highly extended
dust shells around two well known proto-PN,
AFGL 2688 (the Egg nebula), and AFGL 618. In both cases 
the dust emission extends out to a radius of 300-400 \arcsec, 
whilst the previously reported radius of the optical
reflection nebula is at least 10 times smaller. The size of the 
molecular emission region is of the same order of magnitude as that 
observed in the optical. 
The circumstellar shells emit in the far-infrared because the
emission is mainly due to cool dust ($\sim$ 30-50 K). 
In the case of the egg nebula, the far-infrared ISOPHOT emission is matched 
by a uniform extended source of dust with size $\sim$ 2 {\rm pc} and two
denser dust shells at radii $\sim$ 0.87 and $\sim$ 1.7 {\rm pc} 
(with the adopted distance of 1.2 {\rm Kpc}).
The emission of AFGL 618 is matched by a weak extended source with 
radius $\sim$ 3.3 {\rm pc} and two enhancements in emission at radii 
$\sim$ 1.3 and $\sim$ 2.3 {\rm pc} (with the adopted distance of 1.7 {\rm
Kpc}). The sizes of the observed shells are in agreement with the size
predictions of our models (see Table~3).

All of these objects have spherical symmetric shells which implies that no
source of asymmetry is needed in the mass-loss rate process from the AGB, and
confirms our predictions that large CSE should be present around AGB stars.
Probably, the existence of these huge dust shells is related to the
mass-loss suffered by the stars during the TP-AGB.
The formation of structures with sizes as large as $\sim$ 3.3~pc, like those
found around AFGL 618, W Pic, RY Dra or R CrB, is not predicted by any of our
models, however the distance errors and a lower density for the ISM
(those are objects at high Galactic latitude), should be taken into account.

With regards to the emission enhancements in the PPNe AFGL 618 and AFGL 2688,
Speck \etal (2000) related their existence to the occurrence of thermal
pulses. By comparing the dynamical timescales of the shells with the thermal
pulses timescale they derived the mass of the star. This kind of analysis is
widely done in the literature. However,  
the non-linear evolution of the shells generated by the thermal pulses, and
the deceleration introduced by the ISM cause that the models with
different evolutionary
AGB give rise to the formation of structures with similar
sizes. According to our models, does not seem possible to relate the dynamical
timescales derived from CSE with the theoretical timescales associated with
the thermal pulses. 

The only known CSE surrounding
an AGB star which shows multiple shell structures is
IRC+10216. 
IRC+10216, is the CSE of the long period Mira-type
variable carbon star CW Leo has been widely observed at infrared and
millimeter wavelengths. Its distance is estimated to be in the range 120
(Loup \etal 1993) to 170 pc (Winters  
\etal 1994) and is considered to be at the end of its AGB evolution
with a mass-loss rate of $1.5\times10^{-5}$ \mlr (Huggins 1995)
and a terminal 
wind velocity of $\sim$ 14-15 \kms~(Olofsson \etal 1993).
It is the nearest carbon star that has a high mass-loss rate and
consequently is one of the best observed AGB stars.
Recently, deep B and V-band and archival HST WFPC2 images
revealed the existence of an extended circular halo with
multiple shell structures in the extended CSE IRC+10216 (Mauron \& Huggins
2000). The envelope
is detected out to 200 \arcsec. 
The brightness of the shells is explained in terms of the 
illumination of dust-scattered ambient Galactic light.
The observed intensity is proportional to the column density along any line of
sight for an optically thin limit, when the source of illumination is the 
interstellar radiation field.

We have computed the optical intensity from our models
assuming that it is proportional to the column density along the
line of sight. We show two examples that could match the observed
structure found in IRC+10216 (see Fig.~7 from Mauron \& Huggins
2000). The left panel of Fig.~15
corresponds to the 1~\Mso stellar models at 97\% of its AGB evolution and
the right panel to the 1.5~\Mso stellar model at its 93\% of its AGB
evolution. These are the only two models that at the end of their AGB
evolution  
have enough density peaks for account to the ones observed
in IRC+10216 (for a direct comparison see Fig.~15 and Fig.~7 from
Mauron \& Huggins 2000). The central star in both
models suffers from mass-loss rates
consistent with the rates observed in IRC+10216 ($0.35\times10^{-5}$ and 
$0.8\times10^{-5}$ \mlr respectively). For the 1.5~\Mso model the terminal
wind velocity is exactly what is observed, but for the 1~\Mso star model it is
a factor of two lower. By confronting the predictions of
nucleosynthesis models 
with the measurement of isotopic ratios in the CSE of CW Leo, Kahane \etal
(2000) infered a low progenitor mass for the star, M $\le$ 2~M$_{\rm \odot}$. 
In the last few years models have appeared in the literature to
explain the origin of multiple shell events in CSE or post-AGB stars.
Some of them invoke the presence of a binary companion with an eccentric 
orbit (Harpaz, Rappaport \& Soker 1997) or 
a detached binary companion (Mastrodemos \& Morris 1999). All the models 
involving binary companions predicts strict regularity in the
arc spacing between the shells which are not observed. The presence of
a binary interaction together with  
a dynamo mechanism in the AGB star, such as that experienced by the Sun
has been proposed by Soker (2000). 

The shells observed in IRC+10216
shows trends such as increases in shell thickness with increasing radius, and
a wide range of shell spacing, which cannot be accounted by the current
models proposed for the formation of discrete shells in CSE envelopes.
Although, we do not reproduce either the number of
shells or 
the global size of the CSE envelope, the properties of the shells,
such as the thickness increase with increasing radius, 
and the wide range of shell spacing are well characterized by our models.
Evidence for shell 
interactions have been also pointed out by Mauron \& Huggins (2000).

The shells found around IRC+10216 do not show many of the characteristics
found in the shells around PPN and PNe. Moreover,
by the end of the AGB evolution our models predict that the shells
disappear. Therefore, 
if our proposed mechanism for IRC+10216 is correct, it cannot be the same
one that forms the shells observed around PPN. These arcs found around
proto-PN must have a more recent origin (see Garc\'{\i}a-Segura \etal 2001).  

\subsection{The imprints of stellar evolution in the shells}
Since we know the evolution of the wind from the star, our aim 
now is to investigate what are the characteristics that remain
recorded in the final observable structure.

As an example we have made a simple test.
We have chosen the 1~\Mso model at 3.8$\times10^5$ {\rm yr} since the
begining of the evolution. The density structure consists of three well
defined shells. Let's now examine the processes by which these three shells
are 
formed. The outer shell is formed at 1.8$\times10^5$ {\rm yr} and
is continuously fed by the mass-loss from the star during {\it II}.
The intermediate shell is caused by the decrease in
velocity experienced at the end of {\it II}, while the inner
shell is produced later by the decrease in the mass-loss associated
with the end of the same pulse (see Fig.~1).
The difference in time between the formation of the inner and the
intermediate shell is 4,000 {\rm yr} while between 
the intermediate and the outer it is 190,600 {\rm yr}.

Let's now assume that we are able to observe this density
structure and try to infer the
mass-loss modulation which gives rise to its formation.
In principle we should measure shell separations of
0.147 and 0.37 pc respectively.
If we assume a typical outflow speed of about 15 \kms then this gives us 
timescales of 9,590 {\rm yr} and 18,100 {\rm yr} for the formation of the
inner to 
intermediate shell and the intermediate to outer shell respectively.
Therefore, we could conclude that the shells
have been formed by mass-loss modulations separated
by these timescales.
We therefore derive timescales which are more than 2 
times longer for the time interval between the inner and intermediate
shells 
and more than 10 times shorter for the time interval between the
intermediate and outer shells. If we include the gas velocity information
the situation becomes even worse; we derive timescales of 65 and 14,500~{\rm
  yr} respectively. 

We find that velocity gradients play a key role in the
evolution of the shells and that we cannot infer the mass-loss
modulations that give rise to their formation from observations. 
Observations do not provide much information about the mass-loss history.

\subsection{Implications for the stellar evolution models}
Our simulations provide an external shell slightly larger
than what is observed on average. According to our results the size of the
external shell is mainly determined by the AGB 
evolutionary time. If this is the case, shorter AGB evolutionary times than
those obtained by VW93 could connect our models to the existing data. 
It appears that the observed AGB evolutionary time (obtained from HR
diagrams) is shorter than  
that computed by stellar evolution models. The AGB evolutionary time could be
reduced by 
increasing the mass-loss rate during the thermal pulses. 
Moreover, the
maximum mass-loss rate in VW93 models was 
limited to the radiation pressure value which corresponds to the maximum
momentum 
transfer due to simple `scattering' from the electrons to the gas. It is
known that multiple scattering can significantly increase this amount (by
factors from 5 to 10) and it could be that the maximum mass-loss rates are
underestimated. If the increase in
mass-loss rate 
considering multiple scattering does not shorten the AGB evolution
sufficiently, 
other mechanisms together with radiation pressure on dust
grains should be invoked in the mass-loss rate processes (i.e magnetic fields
as proposed by Pascoli 1994). It has also been 
argued that in order to account for the formation of molecular 
species such as titanium carbides observed in CSE around AGB stars (Molster
2000) higher 
mass-loss rate than those predicted by current models should be invoked.
The problem is that observations of the CSE cannot constrain 
the numerical simulations. The difficulties in detecting these huge
circumstellar shells at infrared
wavelengths bias the observations and hence the
constrains we can place on the stellar evolution
calculations using our models. 

\section{Conclusions}
We have studied the response of the circumstellar gas to the large
temporal variations of the wind during the TP-AGB given by VW93 taking into 
account the 
influence of the external ISM. We shown how the mass-loss and velocity
variations 
associated with the thermal pulses lead to the formation of transient
features in the density structure and that these features do not
survive at the end of the TP-AGB evolution.
The evolution of the circumstellar gas during the AGB phase can be said to be
highly  
non-linear due to the hydrodynamical processes which the gas experiences.

The formation
of stable shell structures occurs via two main processes: shocks, between the
shells formed by two consecutive enhancements of the mass-loss, for stellar
masses 1, 1.5, 2 and 2.5~\Mso, and 
for stellar masses 3.5 and 5~\Mso by continuous accumulation of the
material
ejected by the star in the interaction region with the swept up ISM gas. 
 
According to our models a large fraction of the mass of the external
shells  (up to $\sim$ 70 \% for the 1~\Mso stellar model) belongs 
to the ISM matter that had been swept up by the stellar wind. 
This amount decreases according to the assumed density of the ISM, and
according to 
the progenitor mass; the higher the progenitor mass the lower the fraction
of the ISM material residing in the shells.

We quantified the effect of the ISM pressure on the circumstellar gas
structure, and found that the presence of the ISM provides
a non-negligible source of pressure that decelerates and compresses 
the external shells. The velocity of the external 
shell increases by a
factor of $\sim 1.5$ and the density decreases by a factor of $\sim 0.2$ when
the interstellar medium provides ten times less thermal pressure.
 
From our simulations we find that the history of mass-loss history
experienced 
by the star is not recorded in the density or velocity structure of the
gas. Therefore, does not seem possible to use observations to recover
information on the later stages  
of the stellar evolution, such as the 
number of thermal pulses experienced by the star or the 
time between consecutive mass-loss events.
      
According to our simulations, large regions
of neutral gas surrounding the molecular envelopes of AGB stars should be
expected. These large regions 
of gas are formed from the mass-loss experienced by the star 
during the AGB evolution. These large shells should be detectable at infrared
wavelengths.

We thank M. L. Norman and the Laboratory for Computational Astrophysics for
the use of ZEUS-3D. We also want to thank Emanuel Vassiliadis for his fruitful
comments at the begining of this work and for providing us with some of his
models. EV is grateful to Tariq Shahbaz and Letizia Stanghellini for
their careful reading of the manuscript and their valuable comments. The
work of EV and AM is supported by 
the Spanish DGES grant PB97-1435-C02-01. GGS is partially supported by
grants from 
DGAPA-UNAM (IN130698, IN117799 \& IN114199) and CONACyT (32214-E).

\newpage

\begin{deluxetable}{lcccc}
\tabletypesize{\scriptsize}
\tablenum{1}
\tablewidth{0pt}
\tablecaption{Model values}
\tablehead{\multicolumn{1}{c}{Mass}& 
\multicolumn{1}{c}{TP-AGB time} & 
\multicolumn{1}{c}{Time onset SW} & 
\multicolumn{1}{c}{grid size} & 
\multicolumn{1}{c}{cell sizes}\\ 
\multicolumn{1}{c}{[\Mso]} & 
\multicolumn{1}{c}{[$10^4$ {\rm yr}]} & 
\multicolumn{1}{c}{[$10^4$ {\rm yr}]} &
\multicolumn{1}{c}{[{\rm pc}]} &
\multicolumn{1}{c}{[$10^{15}$ {\rm cm}]}}
\startdata
1   &  49.5  &   25.9       &   2        & 6.17\\         
1.5 &  82.9  &   47.8       &   2.5      & 7.72\\
2   & 118.6  &   88.4       &   2.5      & 7.72\\
2.5 & 220.0  &  191.8       &   3        & 9.26\\
3.5 &  42.8  &   21.1       &   3        & 9.26\\ 
5   &  26.2  &    8.7       &   3        & 9.26\\ 
\enddata
\end{deluxetable}

\begin{deluxetable}{lcccc}
\tabletypesize{\scriptsize}
\tablenum{2}
\tablewidth{355pt}
\tablecaption{Shells parameters of the models}
\tablehead{\multicolumn{1}{l}{Label}& 
\multicolumn{1}{c}{Radius} & 
\multicolumn{1}{c}{Density} & 
\multicolumn{1}{c}{Velocity} & 
\multicolumn{1}{c}{Lifetime}\\ 
\multicolumn{1}{c}{} & 
\multicolumn{1}{c}{[pc]} &
\multicolumn{1}{c}{[$10^{-24}$ {\rm $g~cm^{-3}$}]} & 
\multicolumn{1}{c}{[\kms]} &
\multicolumn{1}{c}{[$10^{3}$ {\rm yr}]}}
\startdata
\sidehead{~~1~\Mso}
\tableline
{\it\bf a} & 0.61   & 5.1    &   9.8  & 37.1 \\
{\it\bf b} & 0.22   & 308.0  &  16.8  & 30.5\\
{\it\bf c} & 0.11   & 45.0   &  11.3  & 21.3 \\ 
{\it\bf ab}& 0.96   & 11.2   &  10.4  & $\infty$  \\
           & 1.31   & 6.9    &   7.5  &        \\
           & 1.59   & 4.6    &   5.5  &  \\
           & 1.66   & 4.3    &   5.15 &            \\
{\it\bf d} & 0.29  & 17.6     &  17.0 & 20 \\
           & 0.46   & 7.1     &  17.5 &      \\
{\it\bf e} & 0.09   & 628.0 &   5.9   & 25  \\
           & 0.13   & 135.0 &   5.62 &     \\
{\it\bf x} & 1.19   & 3.6 &   8.4 &  $\infty$\\
           & 1.26   & 4.5 &   9.0  &    \\
\cutinhead{1.5~\Mso}
{\it\bf a}  & 0.70 & 3.02 &  8.44      & 35.7 \\
{\it\bf b}  & 0.28 & 31.2 & 15.22      & 37.4 \\
{\it\bf ab} & 0.99 & 8.8  &  9.94      &$\infty$ \\
            & 1.32 & 6.5 &  7.70       &  \\
            & 1.59 & 4.6 &  5.64    &  \\
            & 1.83 & 3.7 &  4.39    &  \\ 
{\it\bf c}  & 0.39 & 13.6 &  17.00   & 40 \\
{\it\bf d}  & 0.15 & 11.2 &  13.17 & 20 \\
{\it\bf x}  & 1.28 & 4.3 &  10.00 & $\infty$\\
            & 1.58 & 6.5 &  8.72  &   \\
{\it\bf x1} & 0.66 & 2.7 &  19.14   &  \\*
\cutinhead{2~\Mso}
{\it\bf a} &0.27 & 4.0  &  7.04  &  48   \\
           &0.35 & 3.9  &  6.44  &         \\
{\it\bf b} &0.10 & 13.7  &  9.77  & 18    \\
{\it\bf ab}&0.44 & 3.8  &  5.85 &76 \\
           &0.89 & 2.3 &  2.02  &     \\
{\it\bf c} &0.50 & 3.3  &  7.75  &  56   \\
{\it\bf d }&0.13 & 47.1  & 11.87 &30     \\
{\it\bf cd}&0.69 & 5.06  &  7.79  & 88\\
           &1.05 & 3.8 &   5.02  &          \\
           &1.12 & 3.6 &   4.56  &           \\
        &1.21 & 3.2  &   3.97  &    \\
{\it\bf e} &0.32 & 56.3  &  15.69 & 40    \\
        &0.43 & 32.3 &  15.91  &      \\
        &0.60 & 15.3  &  16.15  &    \\
{\it\bf f} &0.16 & 130  &  16.69  &28   \\
        &0.40 & 18.7 &  17.21  &    \\
{\it\bf g} &0.20 & 179  &  14.86   & 44    \\
{\it\bf x} &0.77 & 1.87  &  11.06  &$\infty$\\
        &0.89 & 4.5 &  10.84  &           \\
        &1.03 & 8.6  & 11.49  &           \\
        &1.68 & 10.3 &  9.08  &        \\
\cutinhead{2.5~\Mso}
{\it\bf a} &  0.51 & 3.7 &  5.77  & 71  \\
{\it\bf b} &  0.16 & 13.5 & 11.87     & 69 \\
{\it\bf b1}&  0.07 & 22.9 &  7.20 & 8 \\
{\it\bf ab}&  0.85 & 3.8  &  4.97 & 87 \\
           &  0.97 & 3.4 &  4.16 &       \\
{\it\bf c} &  0.29 & 41.3  & 14.56& 38  \\
{\it\bf d} &  0.37 & 65.2 & 15.44 &  72  \\
{\it\bf x} &  1.09 & 7.13 &  8.49 & $\infty$   \\
        & 1.71 &  11.9  &  9.80 &     \\
\cutinhead{3.5~\Mso}
{\it\bf a}    &0.30 & 5.7 &   8.44 & $\infty$   \\
{\it\bf b}    &0.07 & 94.9 &  10.93& 15  \\
{\it\bf c}    &0.10 & 80.4 &  11.80& 18  \\
{\it\bf d}    &0.25 & 185.0  &  15.11 & 60  \\
{\it\bf ab}   &0.42 & 5.5 &   7.59 & $\infty$  \\
              &0.58 & 4.7 &   5.98 &   \\
{\it\bf abc}  &0.82 & 5.4 &   6.99 &$\infty$ \\
{\it\bf abcd} &1.50 & 18.2&  11.39 & $\infty$ \\
\cutinhead{5~\Mso}
{\it\bf a}& 0.18 &  6.5&  11.40& $\infty$ \\
  & 0.67 &  8.5 &   8.92  & \\
  & 1.22 & 21.0 &  12.16  & \\
  & 1.87 & 16.8 &  11.59 & \\
\enddata
\end{deluxetable}

\begin{deluxetable}{lrrrrrrrrrrrrrrr}
\tabletypesize{\scriptsize}
\tablenum{3}
\tablewidth{0pt}
\tablecaption{CSE properties}
\tablehead{
\colhead{Mass (\Mso)}& 
\multicolumn{3}{c}{Radius [pc]} & \colhead{} &
\multicolumn{3}{c}{$\rho$ ($10^{-24}$ {\rm $g~cm^{-3}$})} & \colhead{}&
\multicolumn{3}{c}{v (\kms)} & \colhead{}&
\multicolumn{3}{c}{Age ($10^{5}$ yr)}\\ 
\cline{2-4} \cline{6-8} \cline{10-12} \cline{14-16}
\colhead{} &
\colhead{{\bf A}} &
\colhead{{\bf B}} &
\colhead{{\bf B/A}} & \colhead{}&
\colhead{{\bf A}} &
\colhead{{\bf B}} &
\colhead{{\bf B/A}} &\colhead{}&
\colhead{{\bf A}} &
\colhead{{\bf B}} &
\colhead{{\bf B/A}} &\colhead{}&
\colhead{{\bf A}} &
\colhead{{\bf B}} &
\colhead{{\bf B/A}}}
\startdata
1  & 1.71 & 2.08 & 1.2  && 4   & 0.9  & 0.2 & & 4.9 & 9.5 & 1.9 &&3.4&2.15& 0.6\\
1.5& 1.84 & 2.5  & 1.4  && 6.8 & 0.6& 0.08 && 7.9   & 7.6 & 1   &&2.3&3.2 &1.4\\
2  & 2.07 & 2.41 & 1.2 && 6.5 & 1.5& 0.2  &&  7    & 12.9& 1.8 &&2.9&1.8 &0.6\\
2.5& 2.2  & 2.58 & 1.2  && 8.3 & 1.4& 0.2  &&  8    & 13.2& 1.7 &&2.7&1.90&0.7\\
3.5& 2.03 & 2.22 & 1.2  && 14  & 2.8& 0.2  && 10.4  & 14.7& 1.4 &&1.9&1.5 &0.8\\
5  & 2.02 & 2.30 & 1.2  && 16  & 2.8& 0.2  && 11.2  & 15  & 1.3 &&1.8&1.5 &0.8\\
\enddata
\end{deluxetable}

\newpage

\begin{figure}
\plotone{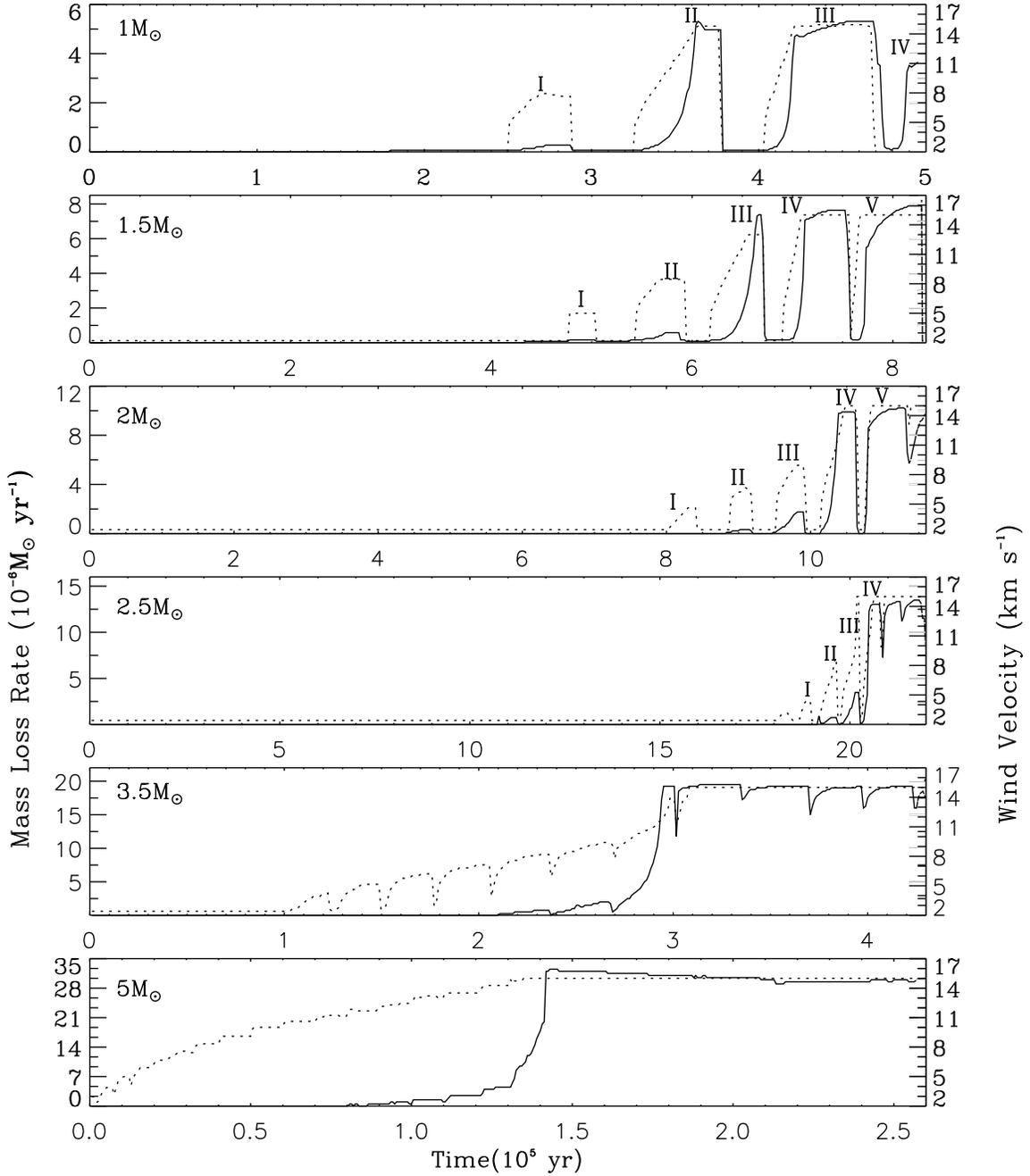} 
\caption[]{
Evolution during the TP-AGB of the mass-loss rate 
    ($10^{-6}~M_{\odot}yr^{-1}$) and wind expansion velocity (\kms) for 
    the 1, 1.5, 2, 2.5, 3.5 and 5~\Mso stellar models.
    The stellar mass is indicated in the top left corner of each
    plot. The data have been taken from 
    Vassiliadis (1992). The solid line shows the mass-loss rate (left scale)
    and the dotted line shows the terminal wind velocity (right scale). Note
    that the 
    lower limits for the mass-loss rate and velocity cannot be appreciated in
    these linear plots. 
\label{f1}}
\end{figure}

\begin{figure}
\plotone{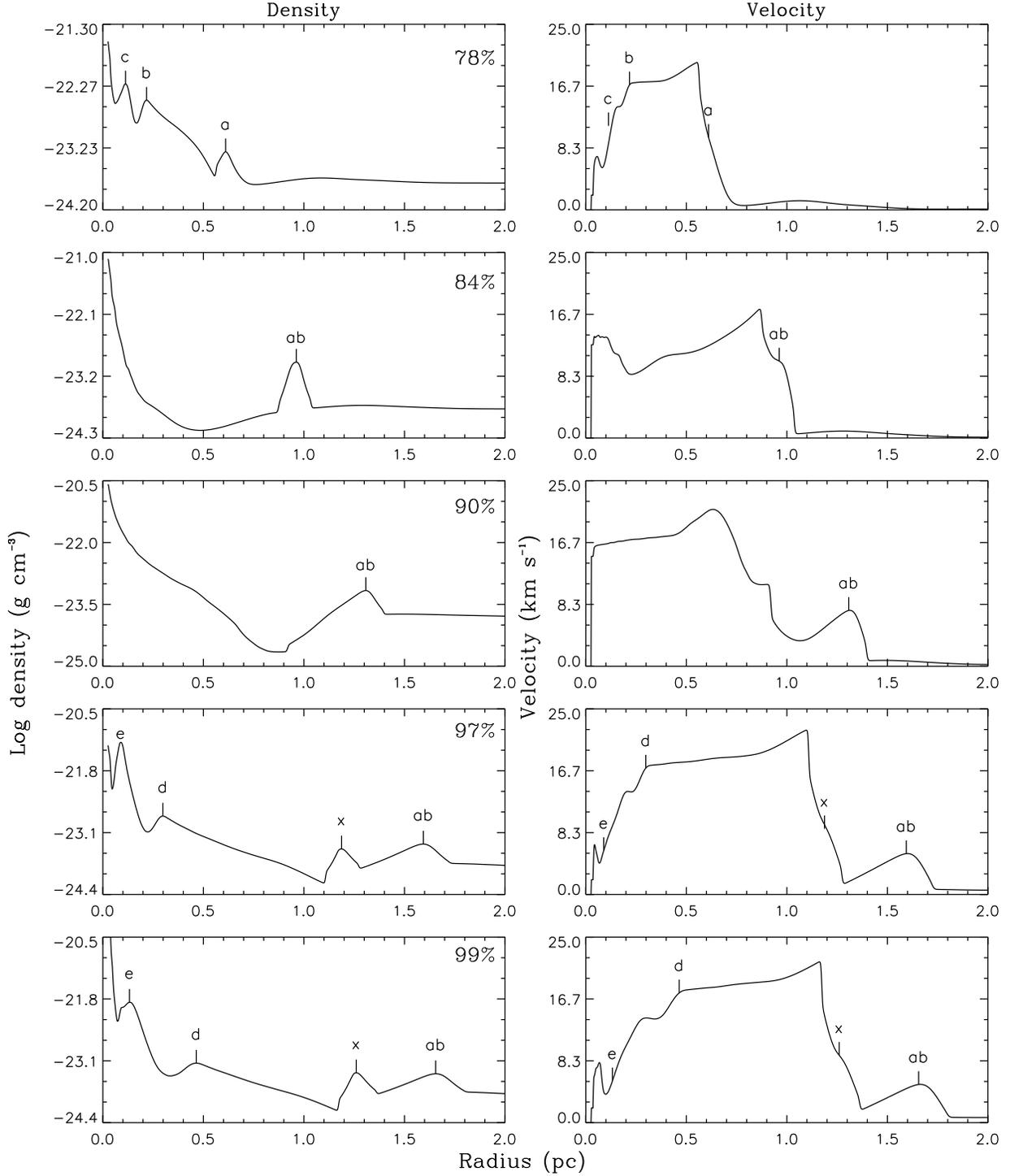} 
\caption[ ]{
Logarithm of the density (${\rm g~cm^{-3}}$) (left column)
    and velocity (\kms) (right column) radial profiles 
    at different times during the AGB evolution for the 1 \Mso stellar
    model. 
    Each density peak is identified and labeled in the density and velocity
    diagrams. In the 
    top right corner of each density plot we give the percentage of the
    total AGB time at which the plots has been selected.
\label{f2}}
\end{figure}

\begin{figure}
\plotone{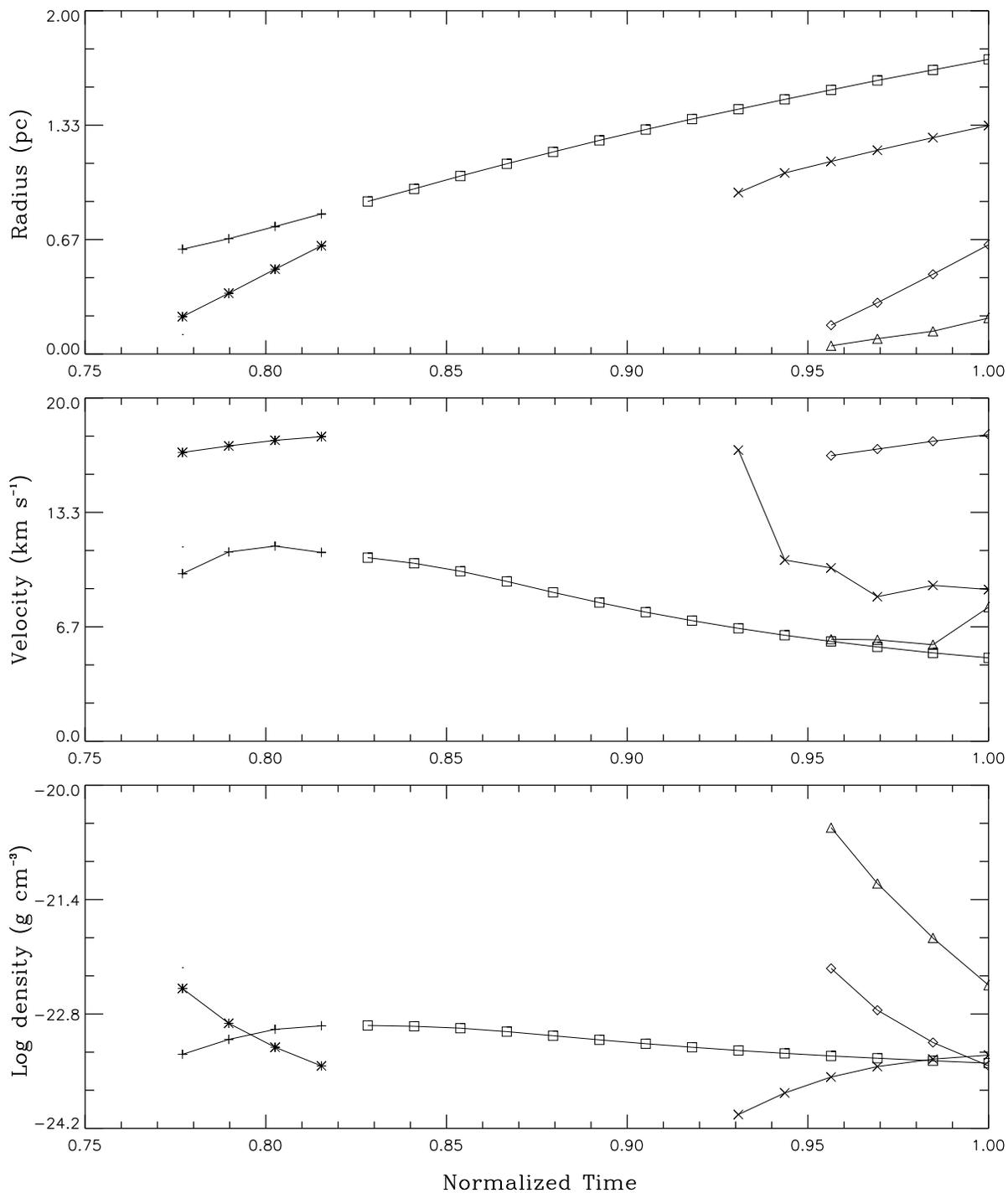} 
\caption[ ]{
Temporal behaviour of the different shells formed
    during the TP-AGB for the 1 \Mso model in terms of radius, velocity and
    density. The time has 
    been normalized to the value at the end of the TP-AGB. The
    plus simbols correspond to the shells {\bf\it a}, asterisks to
    {\bf\it b}, squares to {\bf\it ab}, x to {\bf\it x}, 
    triangles to {\bf\it e} and diamonds to {\bf\it d}.
\label{f3}}
\end{figure}

\begin{figure}
\plotone{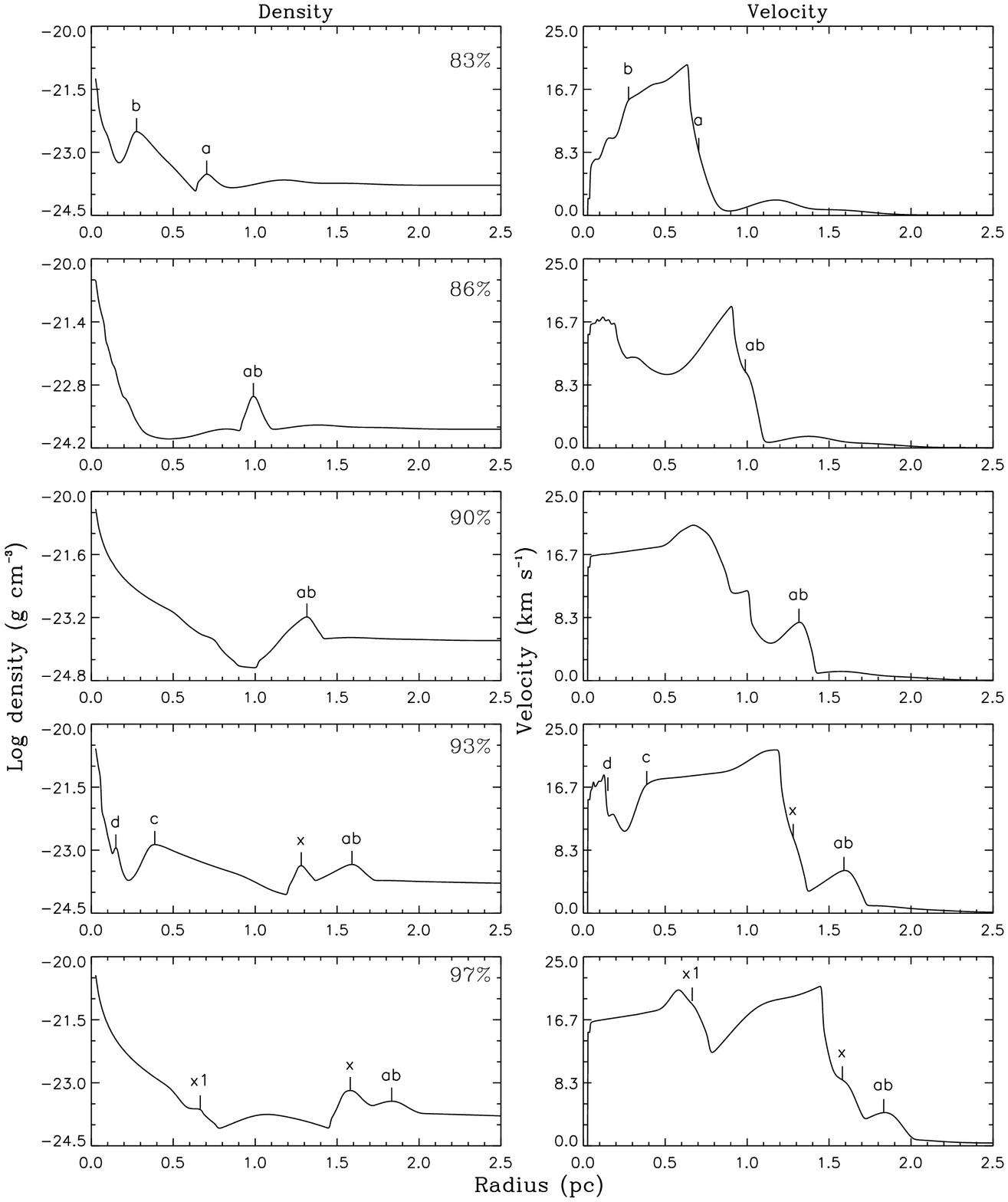} 
\caption[ ]{
The same as Fig.~2
    but for the 1.5~\Mso stellar model.
\label{f4}}
\end{figure}

\begin{figure}
\plotone{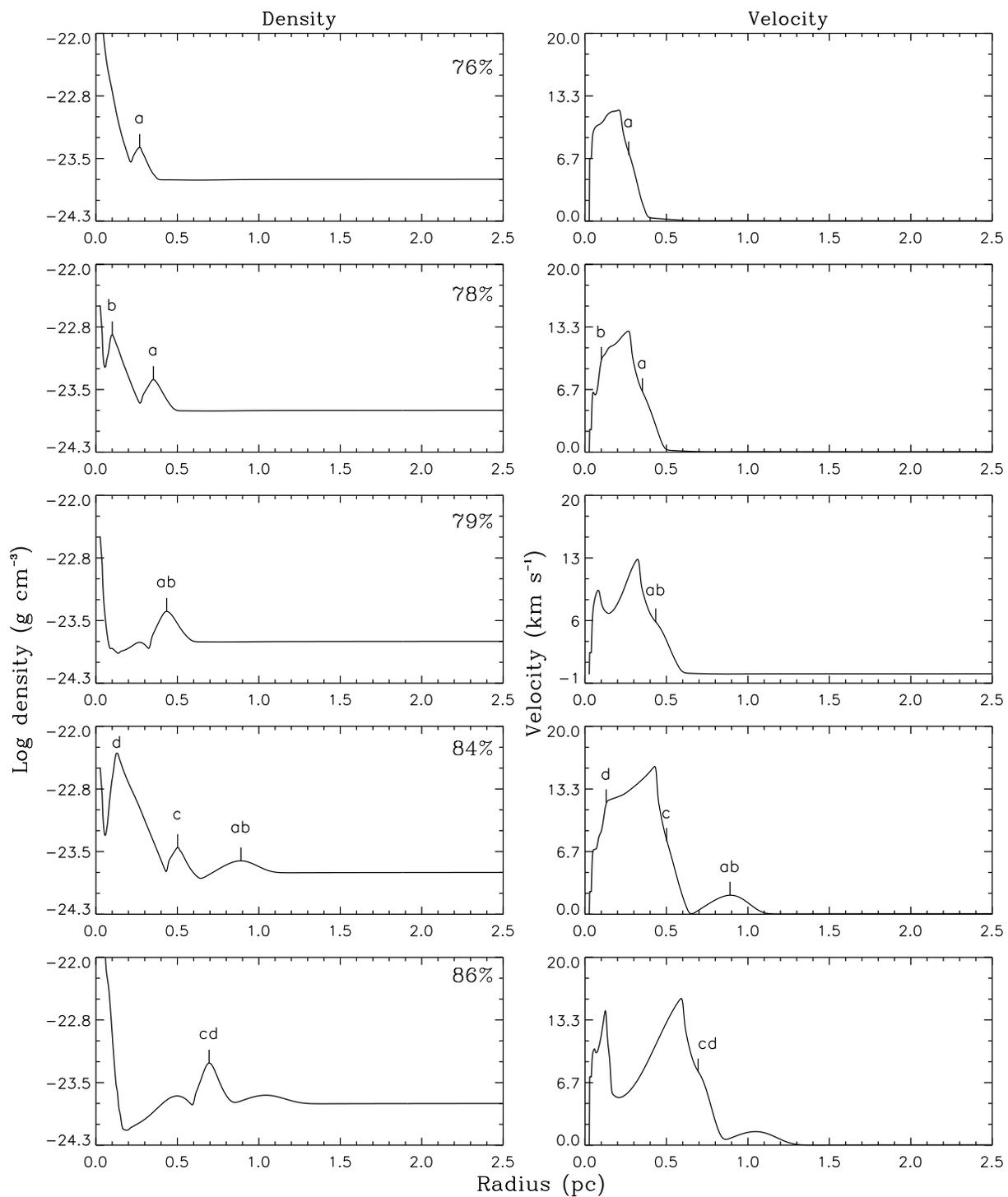} 
\caption[ ]{
The same as Fig.~2
    but for the 2~\Mso stellar model for the first part of the
    evolution. 
\label{f5}}
\end{figure}

\begin{figure}
\plotone{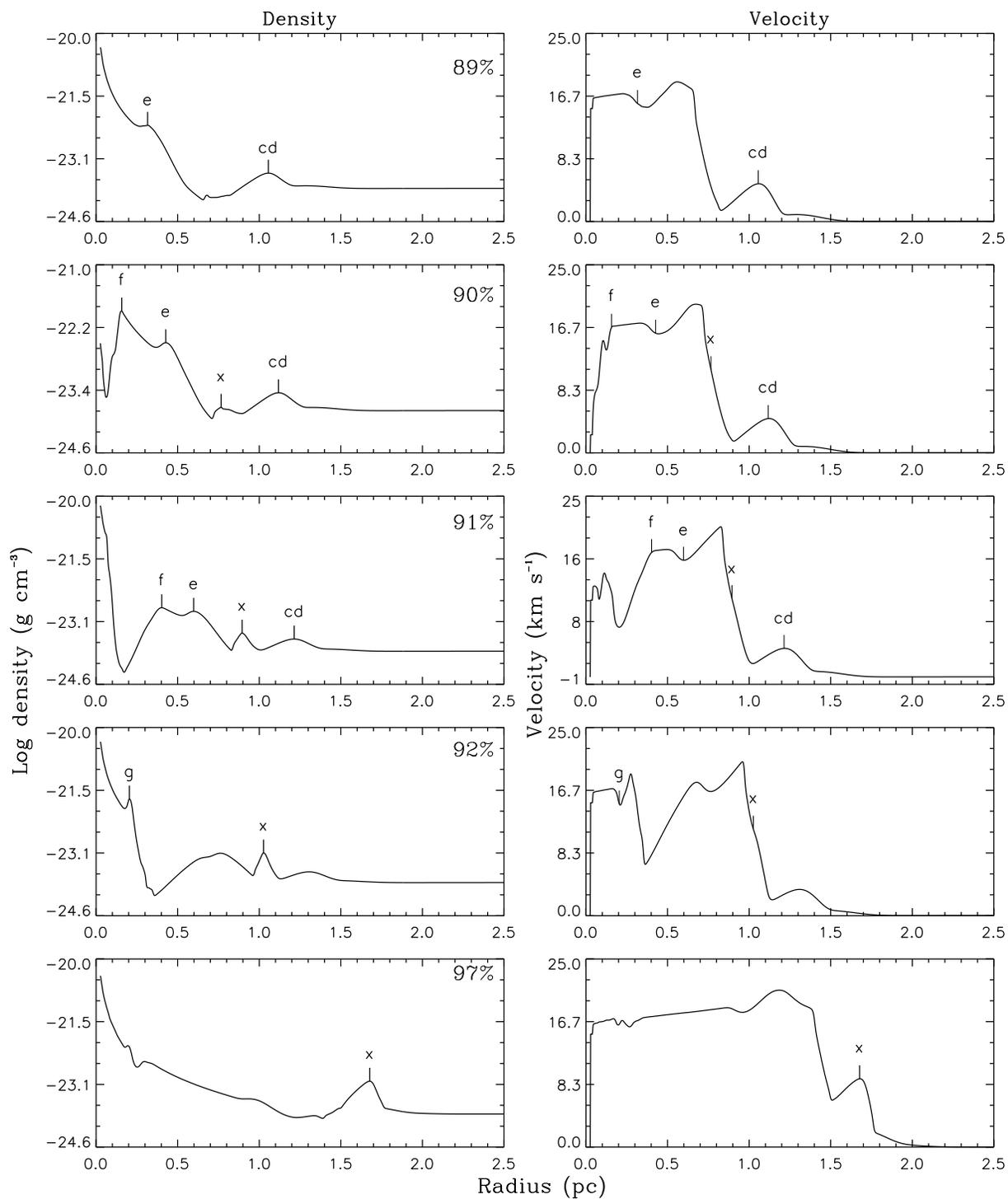} 
\caption[ ]{
The same as Fig.~2
    but for the 2~\Mso stellar model for the second part of the
    evolution. 
\label{f6}}
\end{figure}

\begin{figure}
\plotone{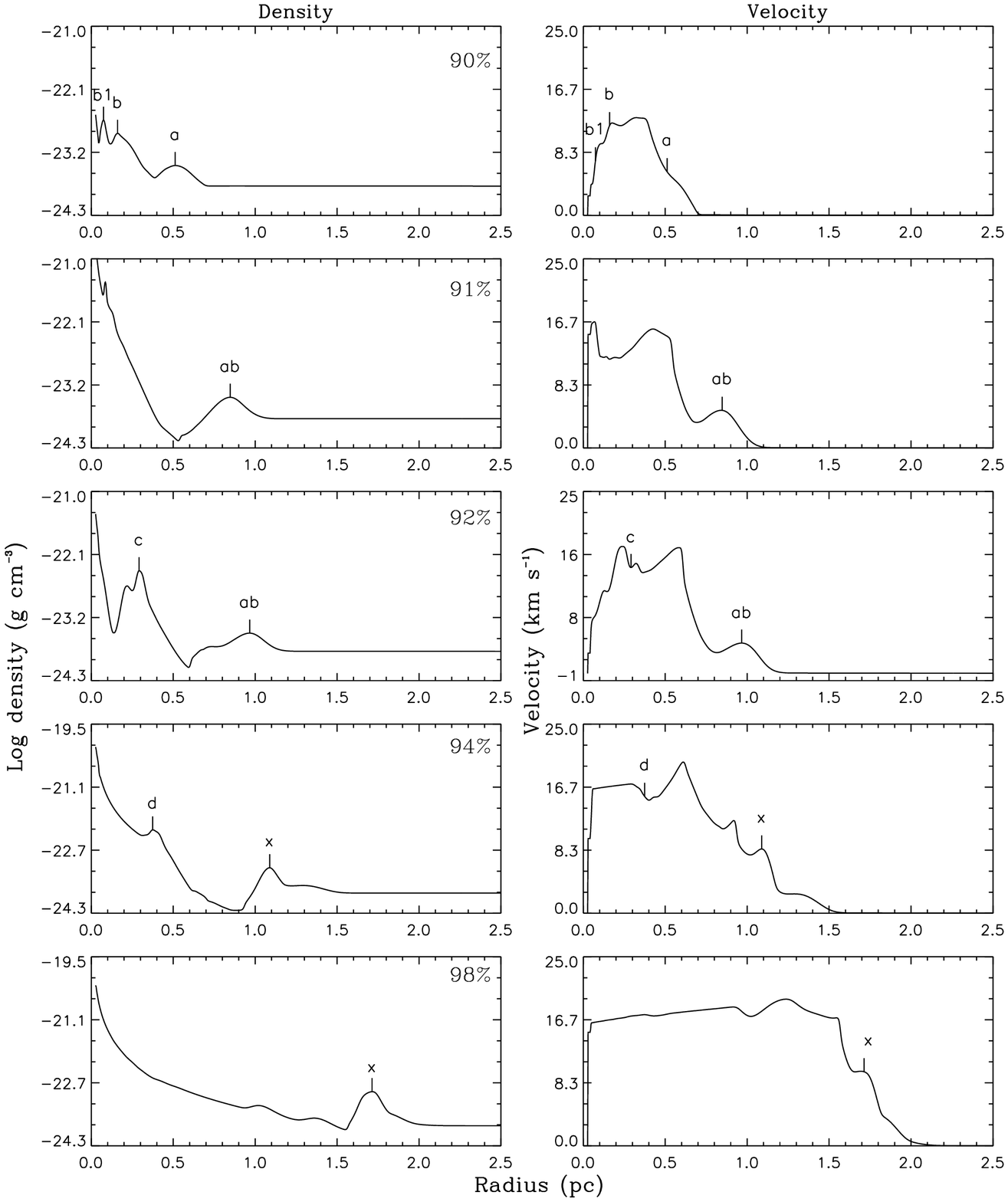} 
\caption[ ]{
The same as Fig.~2
    but for the 2.5~\Mso stellar model.
\label{f7}}
\end{figure}

\begin{figure}
\plotone{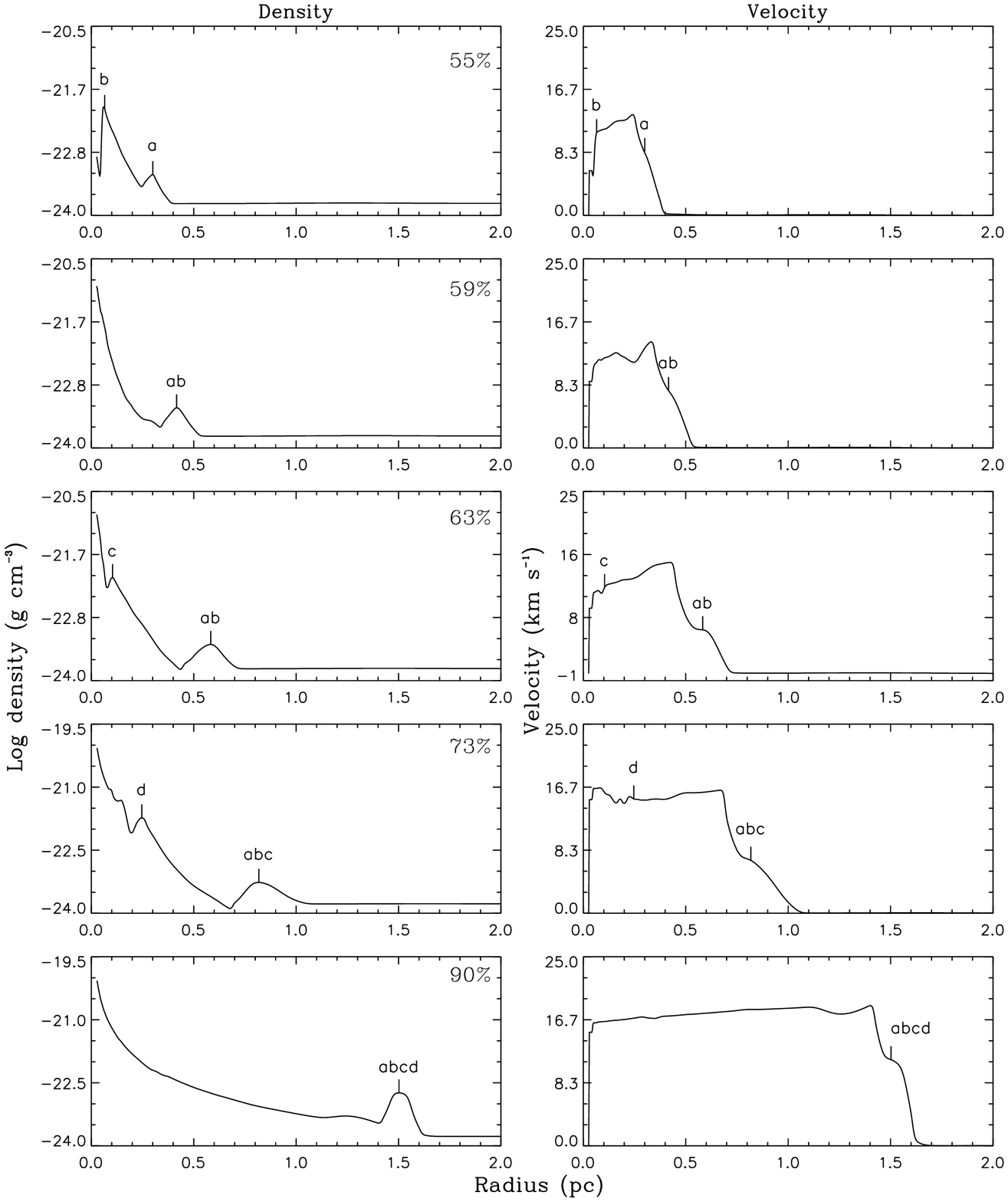} 
\caption[ ]{
The same as Fig.~2
    but for the 3.5~\Mso stellar model.
\label{f8}}
\end{figure}

\begin{figure}
\plotone{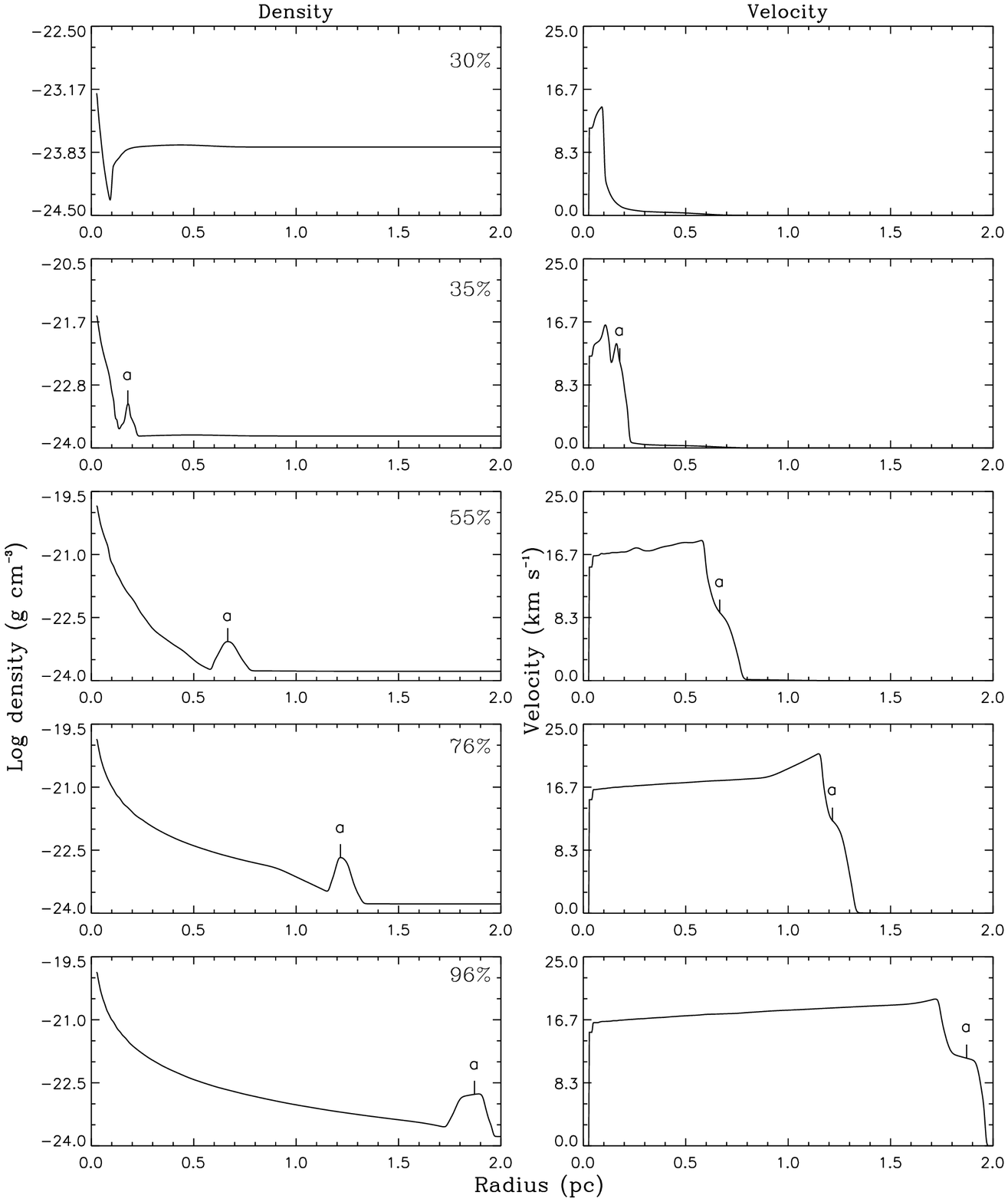} 
\caption[ ]{
The same as Fig.~2
    but for the 5~\Mso stellar model.
\label{f9}}
\end{figure}

\begin{figure}
\plotone{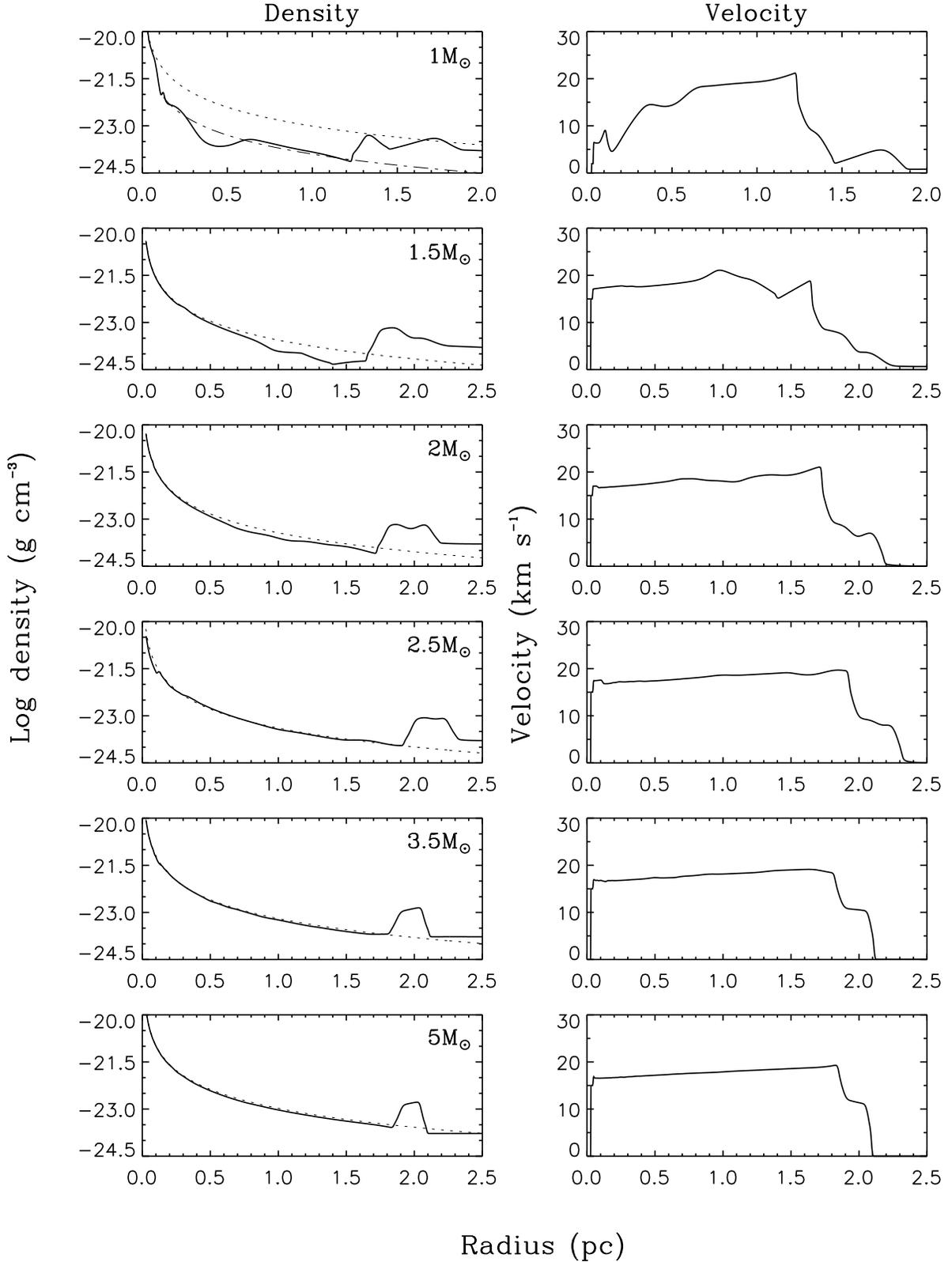} 
\caption[ ]{
Density and velocity structure at the end of the TP-AGB
    for all the stellar models considered in the simulations. The dotted
    line shows the fit to a constant mass-loss rate constant velocity law to
    the density profile. For each stellar
    model the values of mass-loss and velocity for the fits 
    have been taken from the values experienced by the wind at the end of the
    AGB. 
\label{f10}}
\end{figure}

\begin{figure}
\plotone{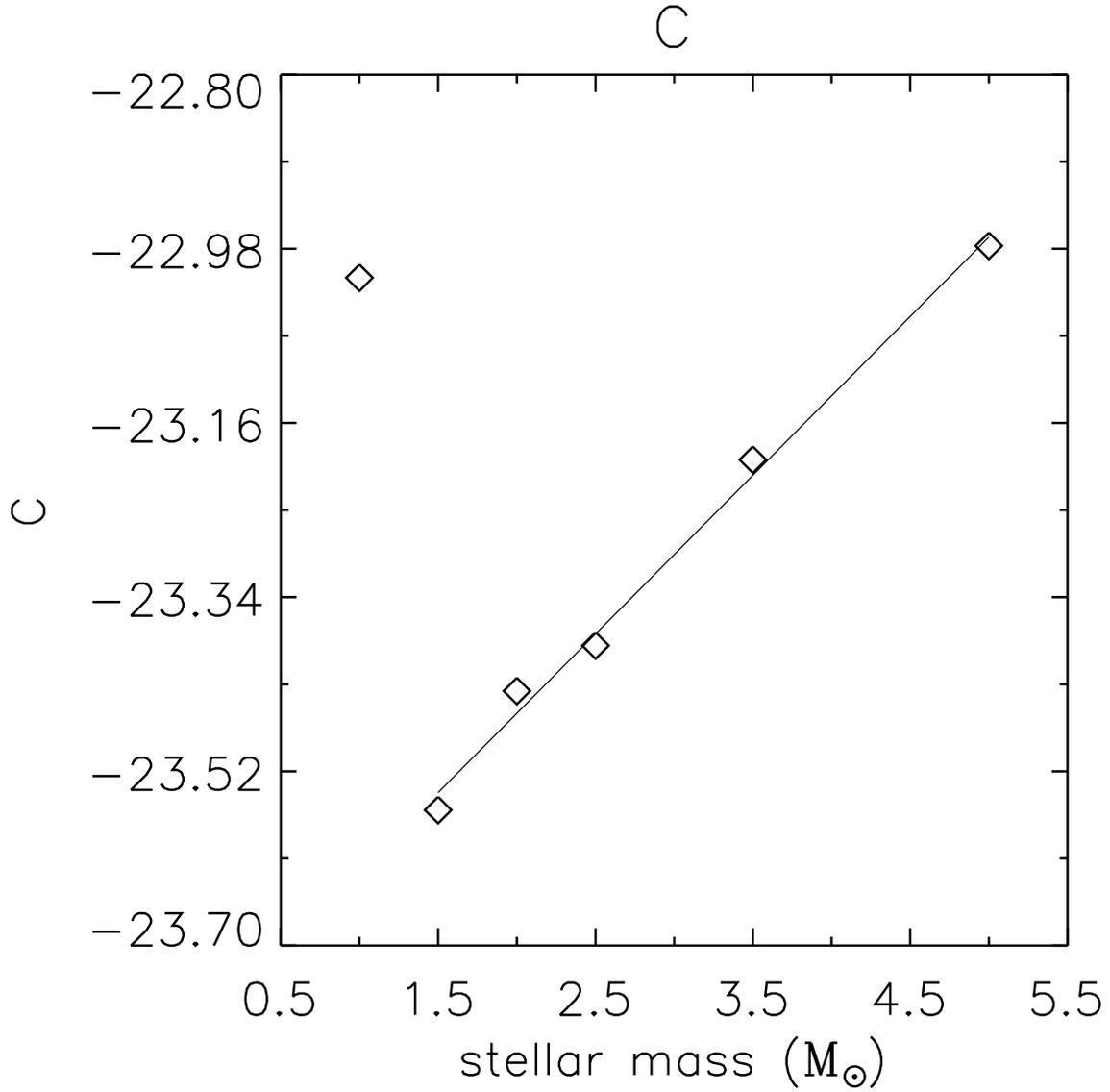} 
\caption[ ]{
The value of the constant $C$ ($C=log(\dot{M}/4~\pi~v_{\rm \infty})$) used to
fit 
the $log \rho=-2~log(r)+C$ law to the density profiles versus 
    stellar mass at the end of the AGB. A linear fit to
    the $C$ values (for M$~>$~1~\Mso) is also shown.
\label{f11}} 
\end{figure}

\begin{figure}
\plotone{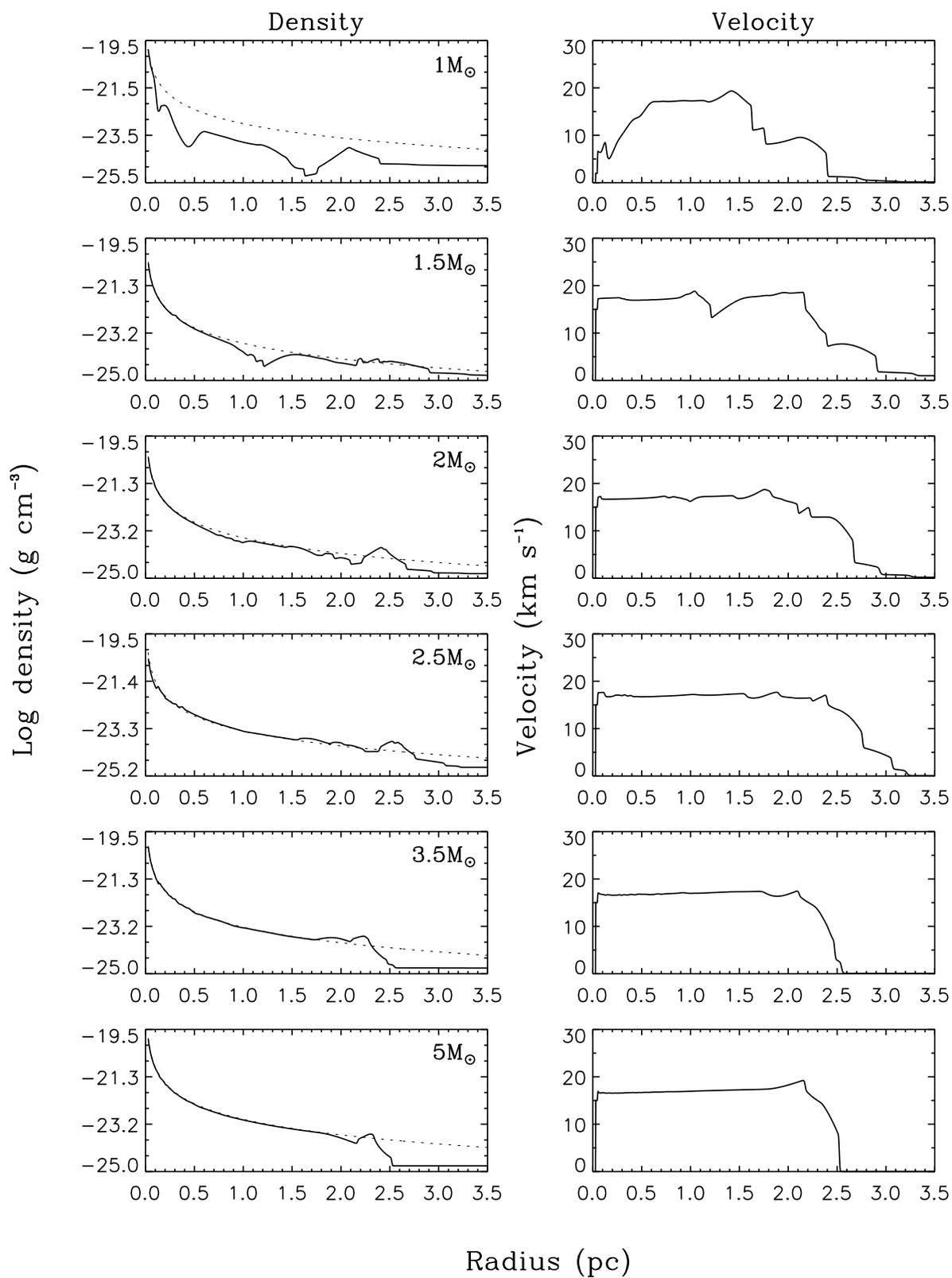} 
\caption[ ]{
Same as fig~10 but for an ISM density
    of 0.1 ${\rm cm^{-3}}$.
\label{f12}}
\end{figure}

\begin{figure}
\plotone{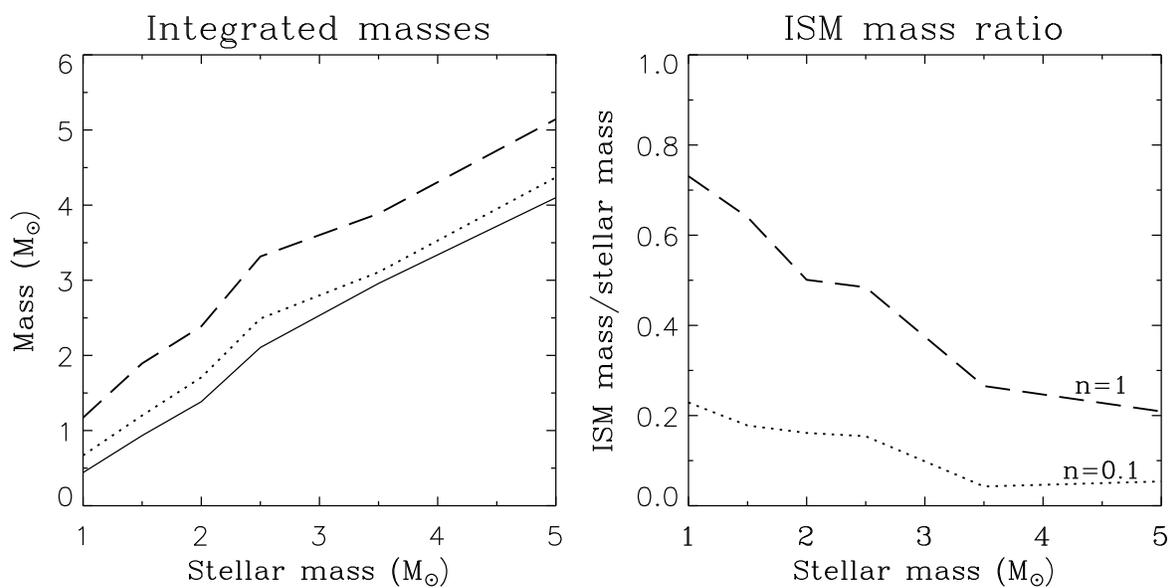} 
\caption[ ]{
Left: Integrated mass above the ISM density 
    at the end of the TP-AGB for each stellar mass. The long-dashed
    and dotted lines
    correspond to ISM densities of 1 and 0.1 ${\rm cm^{-3}}$ respectively. 
    The solid line is the mass lost by the star in solar units. Right: The 
    fraction of the total mass
    belonging to the ISM for each stellar mass. The long-dashed
    and dotted lines
    corresponds to ISM densities of 1 and 0.1 ${\rm cm^{-3}}$ respectively.
\label{f13}}
\end{figure}

\begin{figure}
\plotone{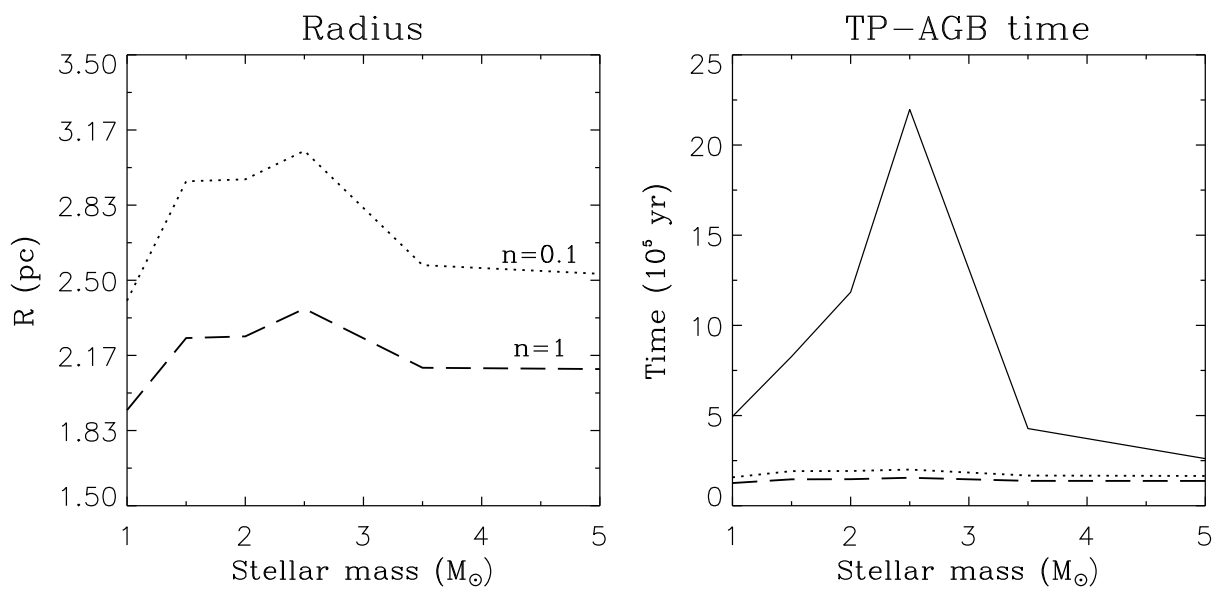}
\caption{
Left: the radius, determined where the external density 
    is higher than the ISM value, versus the stellar mass. 
    The long-dashed and dotted lines
    corresponds to ISM densities of 1 and 0.1 ${\rm cm^{-3}}$
    respectively. Right: the solid 
    black line represents the time the TP-AGB lasts in units of
    $10^{5}$ {\rm yr} for each mass . The long-dashed and dotted lines
    represent 
    the kinematic 
    ages computed assuming a constant flow velocity of 15 \kms for the models
with ISM densities of 1 and 0.1 ${\rm cm^{-3}}$ respectively. 
\label{f14}}
\end{figure}

\begin{figure}
\plotone{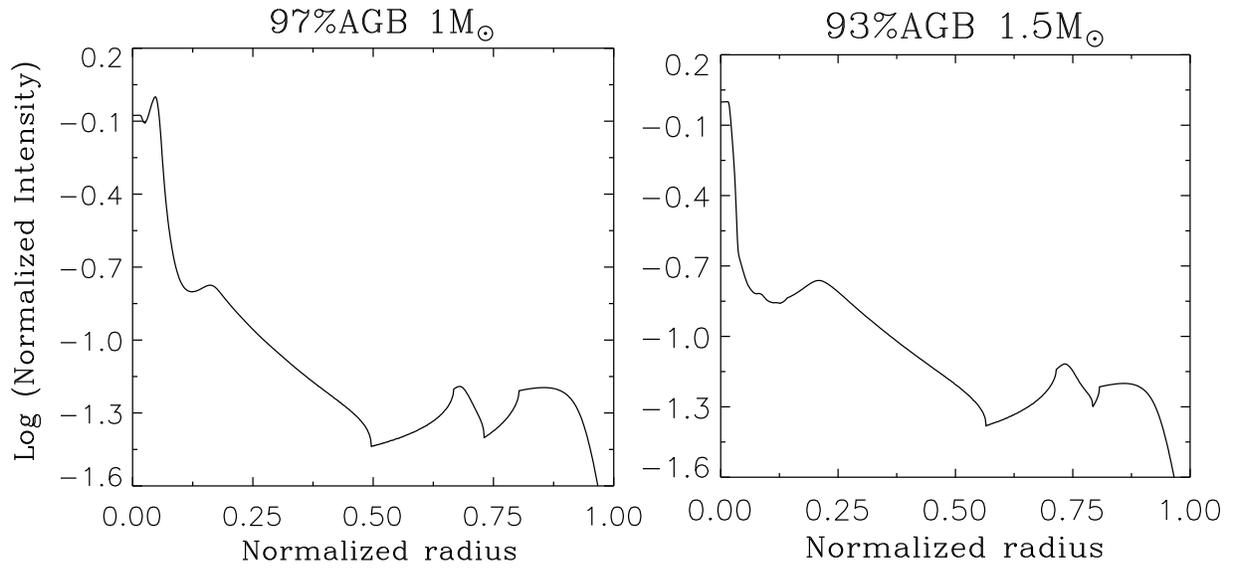}
\figcaption[ ]{
Logarithm of the normalized intensity versus normalized
    radius for the models with 1\Mso (left) and 1.5\Mso (right) at 97\% and
    93 \% of their AGB evolution respectively.
\label{f15}}
\end{figure}

\end{document}